%% file: main.tex
\documentclass[conference,compsoc,letterpaper]{IEEEtran}
\usepackage[nocompress]{cite}

\usepackage{graphicx}
\usepackage{enumitem}
\usepackage{listings}
\usepackage{xcolor}
\usepackage{url}
\usepackage{hyperref}
\hypersetup{hidelinks}
\usepackage{ragged2e}
\usepackage{xspace}
\newcommand\os{operating system\xspace}
\newcommand{\Sname}{\textsc{Lantern}\xspace}
\definecolor{mutationinsert}{RGB}{215,95,0}

\lstdefinelanguage{json}{
  basicstyle=\ttfamily\scriptsize,
  numbers=left,
  numberstyle=\tiny\color{gray},
  showstringspaces=false,
  breaklines=true,
  frame=lines,
  backgroundcolor=\color{gray!5},
  morestring=[b]",
  stringstyle=\color{blue!70!black},
  literate=
   *{:}{{{\color{red}{:}}}}{1}
    {,}{{{\color{red}{,}}}}{1}
    {\{}{{{\color{black}{\{}}}}{1}
    {\}}{{{\color{black}{\}}}}}{1}
    {[}{{{\color{black}{[}}}}{1}
    {]}{{{\color{black}{]}}}}{1}
    {0}{{{\color{orange}0}}}{1}
    {1}{{{\color{orange}1}}}{1}
    {2}{{{\color{orange}2}}}{1}
    {3}{{{\color{orange}3}}}{1}
    {4}{{{\color{orange}4}}}{1}
    {5}{{{\color{orange}5}}}{1}
    {6}{{{\color{orange}6}}}{1}
    {7}{{{\color{orange}7}}}{1}
    {8}{{{\color{orange}8}}}{1}
    {9}{{{\color{orange}9}}}{1}
    {"interfaces"}{{{\color{purple}\bfseries "interfaces"}}}{12}
    {"methods"}{{{\color{purple}\bfseries "methods"}}}{9}
    {"arguments"}{{{\color{purple}\bfseries "arguments"}}}{11}
    {"return_type"}{{{\color{purple}\bfseries "return\_type"}}}{13}
    {"dictionaries"}{{{\color{purple}\bfseries "dictionaries"}}}{14}
    {"fields"}{{{\color{purple}\bfseries "fields"}}}{8}
    {"type"}{{{\color{purple}\bfseries "type"}}}{6}
    {"required"}{{{\color{purple}\bfseries "required"}}}{10}
    {"flags"}{{{\color{purple}\bfseries "flags"}}}{7}
}

\lstdefinestyle{code}{
  basicstyle=\ttfamily\scriptsize,
  numbers=left,
  numberstyle=\tiny\color{gray},
  stepnumber=1,
  numbersep=5pt,
  backgroundcolor=\color{gray!5},
  keywordstyle=\color{blue}\bfseries,
  stringstyle=\color{green!50!black},
  commentstyle=\color{gray}\itshape,
  breaklines=true,
  frame=lines,
  framerule=0.5pt,
  captionpos=b,
  xleftmargin=15pt,
  aboveskip=6pt,
  belowskip=6pt
}

\lstdefinelanguage{JavaScript}{
  keywords={break, case, catch, continue, debugger, default, delete, do, else, false, finally, for, function, if, in, instanceof, new, null, return, switch, this, throw, true, try, typeof, var, void, while, with, const, let, yield, async, await, device, queue, expectValidationError},
  morecomment=[l]{//},
  morecomment=[s]{/*}{*/},
  morestring=[b]',
  morestring=[b]",
  morestring=[s]{`}{`},
  sensitive=true
}

\begin{document}

\title{Dynamic Conformance Testing of WebGPU Through Specification-Driven Mutation}
\author{
    \IEEEauthorblockN{Mahya Samdaliri, Zhihao (Zephyr) Yao, Kasthuri Jayarajah}
    \IEEEauthorblockA{
        \textit{New Jersey Institute of Technology}\\
        \{ms3539, zhihao.yao, kasthuri.jayarajah\}@njit.edu
    }
    }

\maketitle

\begin{abstract}
WebGPU is a low-level graphics and compute API that exposes modern GPU functionality to web applications. While the official WebGPU Conformance Test Suite (CTS) focuses on well-formed usage under the WebGPU specification, it is not designed to stress implementations with semantic edge cases or adversarial inputs. General-purpose fuzzers, in contrast, struggle with WebGPU because of its complex graphics stack and multi-process architecture.
We introduce \Sname, a specification-guided dynamic conformance testing framework that mutates CTS tests using constraints extracted from the WebGPU specification. \Sname extracts explicit syntactic API rules from WebIDL definitions and recovers semantic constraints, such as command ordering and object lifetimes, from natural-language specification text. Selected rules guide AST-located textual transformations that generate both valid and intentionally invalid CTS variants. We execute the resulting tests at scale on AddressSanitizer-instrumented Chromium to discover bugs.
Our evaluation discovers three reproducible bugs, including a heap corruption, in Chromium versions current at the time of study.
These results demonstrate that syntactically seeded, semantics-aware mutation of conformance tests provides a way to uncover browser bugs during WebGPU testing.

\end{abstract}

\begin{IEEEkeywords}
WebGPU, specification-guided mutation, dynamic testing, conformance testing, browser security
\end{IEEEkeywords}

\section{Introduction}
\label{sec:intro}
\begin{figure}[h]
    \centering
    \includegraphics[width=\linewidth]{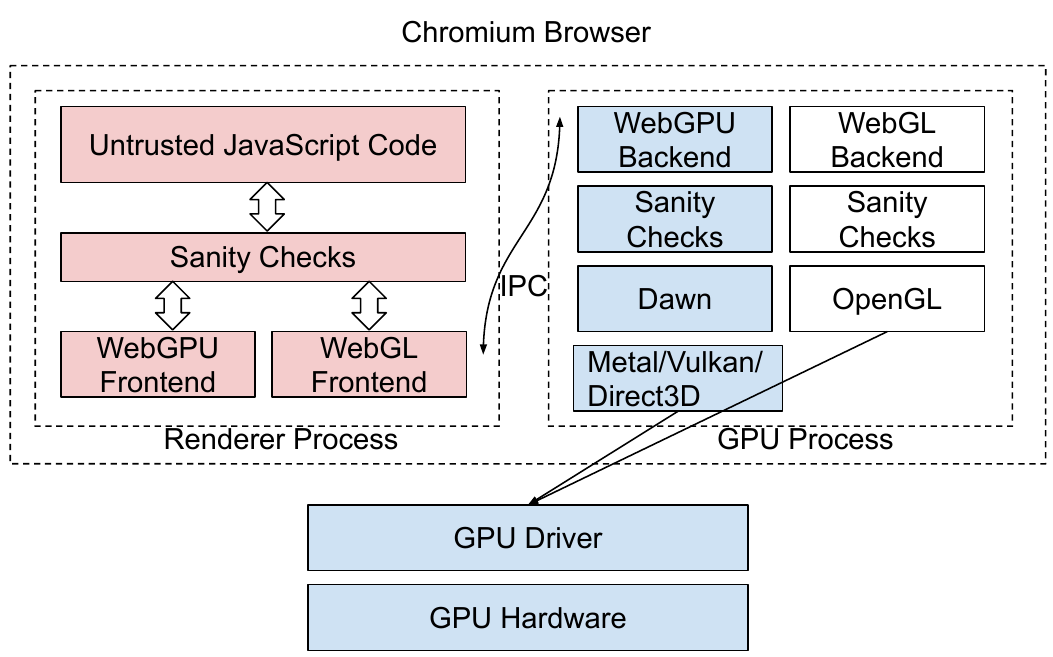}
    \caption{Illustration of the WebGPU and WebGL architecture, including the renderer and GPU processes, \os-specific graphics libraries, and GPU drivers. Dawn is a cross-platform WebGPU layer used in Chromium. Red means untrusted, blue means privileged, and the boxes in gradient show WebGL support.
    }
\label{fig:arch}
\end{figure}

Graphics Processing Unit (GPU)-accelerated graphics in web browsers have revolutionized the visual experience of web applications since the standardization of WebGL, a JavaScript API for rendering 2D and 3D graphics, in 2011 by the Khronos Group~\cite{webgl}.
Its successor, WebGPU, is an emerging web standard that
enables compute shaders and GPU memory management, extending beyond WebGL's fixed-function pipeline and rasterization-only rendering~\cite{webgpu}.
This allows developers to create a wide range of
graphics and computational web applications, such as
in-browser machine learning models such as WebLLM, which performs large language model (LLM) inference entirely within a browser~\cite{webllm}.
The strong performance and flexibility of WebGPU make it an attractive yet still under-studied platform for web developers.
As of this writing, WebGPU remains under active standardization~\cite{webgpu} and browser availability varies across operating systems and GPU configurations~\cite{webgpu_impl_status, chrome_webgpu_status}.
Browsers translate sandboxed WebGPU JavaScript commands to native graphics APIs through a multi-process graphics stack spanning the renderer, a privileged GPU process, and the underlying driver (Figure~\ref{fig:arch}).

However, by giving web applications closer access to GPU hardware, WebGPU inherits and amplifies security concerns familiar from WebGL~\cite{webglsec, webgpu_chrome_report}.
Prior work demonstrates that web-exposed GPU interfaces can leak information, escalate privilege, or enable side channels on the shared system graphics stack
~\cite{yao2018sugar,naghibijouybari2018rendered}.
WebGPU further increases the complexity of validation and state management in memory object management, synchronization, and argument types~\cite{webgpu_chrome_report}, heightening the need for thorough security evaluation.
These properties call for systematic security testing
of WebGPU implementations at development time.

Existing WebGPU testing infrastructure primarily targets conformance:
the official testing framework, the WebGPU Conformance Test Suite (CTS)~\cite{webgpu_cts_repo}, consists of deterministic tests that verify that implementations behave correctly.
While CTS is indispensable for validating specification compliance (in terms of WebGPU support), it is not designed to stress implementations with unexpected or adversarial inputs.
General-purpose fuzzers can generate diverse inputs, but they struggle to meaningfully exercise WebGPU. Unstructured mutations typically violate basic preconditions (e.g., required usage flags, object lifetimes, or call ordering) and terminate during validation, preventing deeper execution~\cite{bugzilla_webgpu_fuzz, wong2025webglitch, simols2024webgpu}. Coverage-guided fuzzing is also difficult because feedback would need to span multiple processes and layers, including the GPU process and the kernel driver. For WebGL, prior work therefore used error messages as a proxy for behavioral feedback~\cite{peng2023gleefuzz}.
A similar challenge persists for WebGPU: existing WebGPU fuzzers either generate randomized
WebGPU programs (e.g., WG-Fuzz~\cite{simols2024_wg_fuzz, simols2024webgpu}) or use manually curated models (e.g., WebGlitch~\cite{wong2025webglitch}), making it difficult to capture deep dependencies. For example, one CTS-derived test variant in our campaign triggered a security bug in Blink's font cache through a call stack 118 frames deep (\S\ref{subsec:motivation:font-cache}).

To bridge this gap, we introduce \Sname,
a \textbf{\textit{framework for specification-guided dynamic conformance testing of WebGPU}}.
\Sname starts from the official WebGPU CTS and systematically mutates existing tests to explore additional, specification-defined behaviors.
The key idea is to treat the WebGPU specification as a source of test-generation constraints: (i) \emph{explicit syntactic rules} extracted from WebIDL (types, interfaces, and flags), and (ii) \emph{implicit semantic rules} recovered from natural-language requirements (e.g., ordering constraints, object lifetimes, and alignment restrictions). We use GPT-5, a large language model, to assist in identifying candidate implicit rules, and the results are manually validated before use.

The extracted rules are applied to guide Abstract Syntax Tree (AST)-level mutations to JavaScript CTS tests.
\Sname uses the parsed AST to locate object-literal mutation candidates and applies textual edits guided by selected specification-derived constraints.
A configuration parameter controls the mutation \emph{scale} (the probability applied at each rule check) and a \emph{mode} that directs supported handlers toward satisfying (valid) or violating (invalid) a selected local constraint.
\S\ref{sec:eval:mode} discusses how different mutation configurations affect bug discovery.
We execute the mutated suites on ASan-instrumented Chromium builds and collect crash/sanitizer reports for offline triage.

In summary, our main contributions are:
\begin{itemize}[leftmargin=*, nosep]
    \item We introduce \Sname
, a dynamic conformance testing framework for WebGPU that systematically extends the official CTS by mutating existing tests according to specification-derived rules.
    \item We extract explicit API constraints from WebIDL and recover additional behavioral rules from the specification text, using LLM assistance to accelerate rule identification as a preprocessing step.
    \item We design a mutation engine that uses a JavaScript AST to locate candidate object literals and applies selected rules under two intended-direction modes.
    \item We evaluate the resulting test suites on Chromium and identify multiple vulnerabilities.
\end{itemize}

We have responsibly reported three reproducible bug findings to the Google Chromium team.
The V8 memory-safety finding was already known internally to the vendor, but our report supplied an additional triggering test case.

\section{Background}

\subsection{Web-based GPU Acceleration}
The evolution of GPU acceleration in web browsers began with the introduction of WebGL by the Khronos Group in 2011~\cite{webgl}.
WebGL brought GPU-accelerated graphics to the web by exposing a JavaScript API for 2D and 3D rendering. Although widely adopted, WebGL’s limited compute capabilities and largely implicit resource management restrict its flexibility for modern GPU workloads such as Artificial Intelligence (AI) inference.

To overcome these limitations, in 2021, the World Wide Web Consortium (W3C) introduced WebGPU~\cite{webgpu}, a low-level graphics and compute JavaScript API that serves as the successor to WebGL. WebGPU exposes an API closer to \os graphics stacks (e.g.,
Vulkan, Metal, and Direct3D) with compute shaders, command buffers, and explicit resource management.
These features support high-performance GPU use cases,
including real-time rendering, virtual/augmented reality (VR/AR), and machine learning (ML) in the browser~\cite{webllm}.
At the same time, the richer programmability and low-level GPU features enabled by WebGPU expand the browser's attack surface and introduce new challenges for secure API design.

\subsection{Security of WebGL and WebGPU}
Prior work has identified security risks in web-exposed GPU interfaces, including timing side channels, cache-based leakage, and device fingerprinting~\cite{naghibijouybari2018rendered,giner2024generic,ferguson2024webgpu,laor2025poster}. These attacks exploit shared GPU resources and the complexity of the graphics stack to infer sensitive information or system behavior.

WebGPU further increases this risk by exposing lower-level GPU functionality and introducing more complex validation and state-management requirements, which expand the browser attack surface. These observations motivate the need for systematic and automated testing approaches.

\subsection{WebGPU Conformance Test Suite (CTS)}
WebGPU's predecessor, WebGL, has long relied on CTS to ensure cross-platform implementation consistency~\cite{khronos_webgl_cts}. WebGPU inherits this model but introduces greater complexity due to its richer computational API surface, programmable pipelines, and explicit memory management.
The WebGPU CTS~\cite{webgpu_cts_repo} is the official framework for verifying that browser implementations correctly follow the specification. It consists of thousands of deterministic tests that validate structural and behavioral requirements such as adapter queries, buffer usage flags, and render-pass semantics.
Google notes that ``robust WebGPU CTS was invaluable'' for identifying both engine-level implementation issues and underlying driver bugs, and that the challenge is to ``ensure that every possible code path is reliable''~\cite{jones2024webgpu}.

CTS cases are typically written in TypeScript and follow a parameterized style that enumerates specification-defined conditions. While the suite emphasizes \emph{correctness} of API usage, many tests also validate failure behavior by checking that invalid states are rejected with appropriate validation errors. For example, the test shown in Listing~\ref{lst:cts-example} checks that \texttt{writeBuffer()} rejects destroyed or invalid buffers~\cite{github_cts_writebuffer_example}.
Such tests motivate our approach: extending the static CTS into a \emph{dynamic} testing framework to explore the boundaries of specification-defined behavior. We emphasize that our goal is not to replace CTS, but to transform it into a dynamic execution harness for broader and deeper WebGPU evaluation.

\begin{lstlisting}[style=code, language=JavaScript, caption={CTS test case for use-after-destroy in \texttt{writeBuffer()}~\cite{github_cts_writebuffer_example}.}, captionpos=b, label={lst:cts-example}]
t.expectValidationError(() => {
  device.queue.writeBuffer(buffer, 0, data, 0, data.length);
}, bufferState !== 'valid');
\end{lstlisting}

\subsection{AI-Assisted Specification Understanding}
Recent work uses LLMs to generate mutation operators, solve program constraints, and recover protocol structure~\cite{ou2024mutators, yang2025hybrid, meng2024large, rahman2024cellularlint}. This is directly relevant to WebGPU, whose behavioral constraints are split across WebIDL, prose validation rules, and lifecycle descriptions rather than a single formal grammar.

Complementing this line of work, \Sname uses LLM assistance to recover candidate WebGPU behavioral rules that guide AST-located textual mutations.

\section{Motivation and Design Goals}
\label{sec:motivation}
\textbf{Security modeling.} We consider remote attacks in which an adversary controls untrusted WebGPU content, such as a webpage or embedded frame, running in the renderer process. Side channels and attacks requiring local or physical access are out of scope.

The official WebGPU CTS is indispensable for standardization, but it is designed for conformance rather than adversarial testing.
Even though CTS tests include erroneous scenarios (e.g., Listing~\ref{lst:cts-example}),
CTS cases validate a narrowly scoped requirement under a static, scoped execution pattern, and they do not systematically explore longer sequences that combine multiple objects, state transitions, and ordering constraints.
In practice, WebGPU vulnerabilities arise from complex call chains rather than from isolated API misuse~\cite{giner2024generic, levine2025saferace}.

Consider the CTS check in Listing~\ref{lst:cts-example}. A semantics-guided mutator can insert API calls that transition objects between specification-defined states. Listing~\ref{lst:writebuffer-mutation} shows a simplified example in which we insert a state-changing call immediately before \texttt{writeBuffer()}, producing a targeted use-after-destroy pattern.
Extending CTS by hand to cover these interactions does not scale and must be repeated as the specification evolves.

\begin{lstlisting}[style=code, language=Java, caption={\protect\RaggedRight
Semantics-guided mutation of a CTS check for \texttt{writeBuffer()}. Highlighted code is inserted by us for illustration.}, label={lst:writebuffer-mutation}, captionpos=b, escapeinside={(*@}{@*)}]
t.expectValidationError(() => {
  (*@\textcolor{mutationinsert}{\ttfamily\small\bfseries buffer.destroy(); // inserted}@*)
  device.queue.writeBuffer(buffer, 0, data, 0, data.length);
}, true);
\end{lstlisting}

General-purpose fuzzing is also a poor fit for WebGPU, as discussed in \S\ref{sec:intro}. Unstructured mutations commonly violate API preconditions (e.g., required usage flags or call ordering), causing tests to fail during validation and preventing deeper execution. Coverage-guided fuzzing further faces practical barriers in multi-process browser architectures, where feedback must span renderer and GPU processes and is easily dominated by noise in shared subsystems~\cite{peng2023gleefuzz}.

\subsection{Motivating Example: Deep Call Chains in Skia}
\label{subsec:motivation:font-cache}
During our campaign, \Sname triggered a DCHECK in Skia (a component related to CPU-based rendering
) on both Chromium~133 and~146 while running mutated WebGPU CTS tests. The crash occurs in the renderer process after the mutated CTS test drives a long chain of execution through V8 bindings and DOM style/layout before reaching Skia font selection; the renderer stack trace contains over 118 frames.
The immediate reason is a missing \texttt{font \_platform \_data}, but to reach it requires keeping the input executable through many state-dependent steps.
We reported this issue to Google.

This example highlights a complementary testing gap for program-generation fuzzers such as WG-Fuzz~\cite{simols2024_wg_fuzz} and WebGlitch~\cite{wong2025webglitch}: deep browser paths can depend on long sequences of valid API interactions already encoded in CTS tests.

We address this gap with dynamic conformance testing: starting from CTS programs and applying specification-grounded mutations to explore additional behaviors and boundary conditions.
\Sname aims to answer the following research questions:
\begin{itemize}[leftmargin=*, nosep]
\item \textbf{Testing effectiveness.} Can specification-grounded mutation of CTS tests uncover vulnerabilities by exercising behavior beyond the unmodified CTS, while remaining executable in the CTS harness?
  \item \textbf{Mutation strategy.} How do mutation strategy choices, e.g., valid versus invalid mode, mutation scale, parallel or not, affect behavioral diversity and discovery rate? Large-scale fuzzing campaigns typically run many browser instances concurrently to maximize throughput, but WebGPU testing places competing demands on shared GPU and CPU resources; we therefore treat parallelism as an open question rather than assuming it improves coverage, and revisit it directly in \S\ref{subsec:parallel}.
  \item \textbf{Cross-version behavior.} How did Chromium’s WebGPU robustness evolve from version~133 to~146, as reflected in \Sname’s cross-version experiments (\S\ref{subsec:version-comparison})?
\end{itemize}

To answer these questions, we have four design goals:

\noindent\underline{Preserve CTS harness structure.} Mutate object literals within existing CTS tests while reusing their surrounding setup.
The resulting variants are used for fuzzing inputs.

\noindent\underline{Ground mutations in the specification.} Derive mutation constraints from WebIDL and recover additional behavioral rules from the specification prose, reducing reliance on manually curated schemas.

\noindent\underline{Support multiple mutation strategies.} Make the intended mutation direction (valid or invalid mode) and scale (mutation intensity) configurable so we can identify effective testing strategies.

\noindent\underline{End-to-end testing and triage.} Besides running within the CTS harness, \Sname supports monitored runs for sanitizer-instrumented browsers. The mutator accepts a fixed random seed, and generated inputs and logs can be retained
for replay and offline analysis.

\section{Design}
\label{sec:design}

This section describes \Sname's architecture and explains how we extract constraints and behavioral rules to drive each stage of the testing pipeline.

\subsection{System Overview}

\begin{figure}[t]
    \centering
    \includegraphics[width=0.75\linewidth]{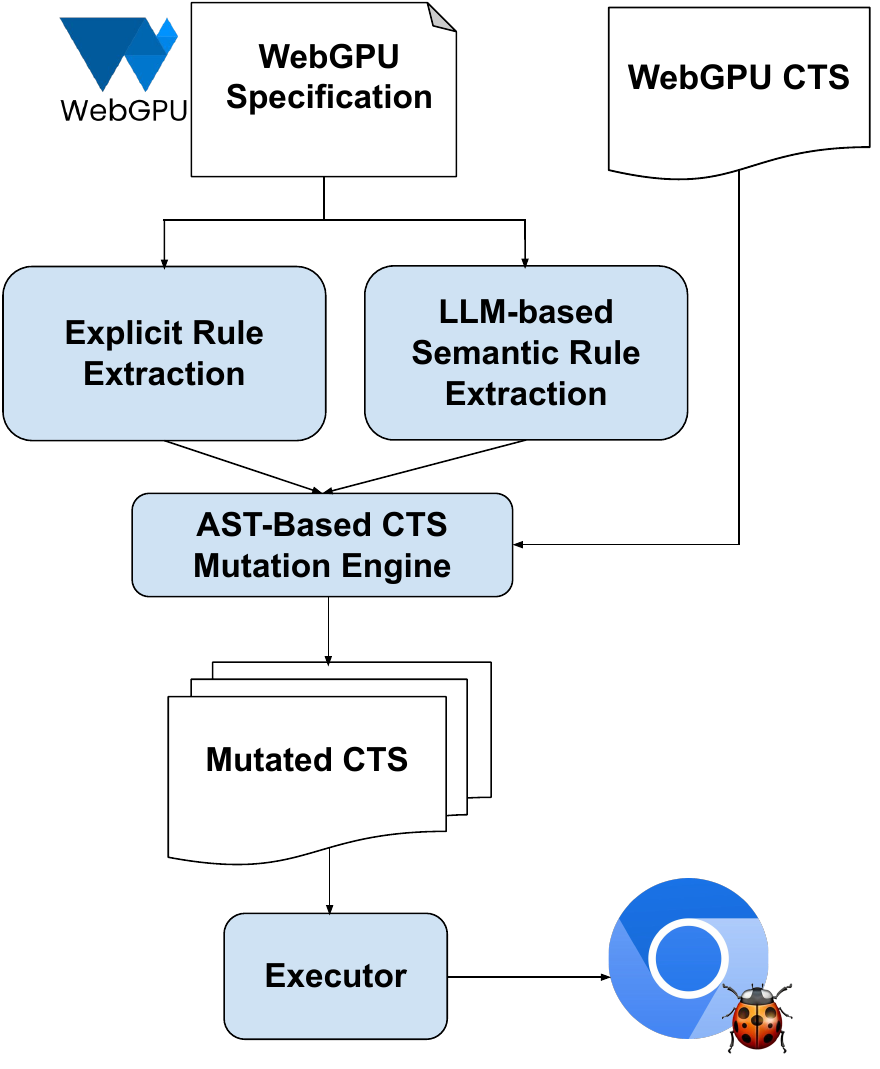}
   \caption{Overview of \Sname, a specification-guided mutation-based framework for dynamic WebGPU conformance testing.
}
    \label{fig:overview}
\end{figure}

Figure~\ref{fig:overview} illustrates the architecture of \Sname, which takes as input (1) the WebGPU specification and (2) the WebGPU CTS corpus, and outputs mutated CTS suites together with replay artifacts for any failures observed during execution. \Sname consists of four stages:

\begin{itemize}[leftmargin=*, nosep]
  \item \textbf{Rule Extraction.}
  Parse WebIDL blocks to obtain explicit typing and structural constraints, and recover additional semantic constraints from the specification prose (e.g., ordering, lifetimes, and alignment).

  \item \textbf{CTS Mutation.}
  Apply rule-guided AST transformations to CTS JavaScript tests under a configurable mode (valid/invalid) and scale, producing mutated suites that remain runnable in the CTS harness.

  \item \textbf{Execution and bug triage.}
  Run mutated suites on an instrumented browser and collect logs and sanitizer/assertion reports from the renderer and GPU processes.

  \item \textbf{Crash Detection and Analysis.}
  Record mutated inputs, configurations, and browser logs for crash analysis and statistics used to answer the research questions in \S\ref{sec:motivation}.
\end{itemize}

A core design decision is to keep mutation and execution decoupled: the mutator produces a static mutated suite, and the executor can replay it independently. This separation avoids relying on the unavailable online coverage feedback in a multi-process graphics stack, as discussed in \S\ref{sec:intro} and~\cite{peng2023gleefuzz}.

\subsection{Explicit Rule Extraction}
\label{sec:design:explicit}

\Sname first extracts explicit type and interface information from the W3C WebGPU specification, defining the syntactic correctness boundary within which subsequent test mutations must operate. The specification is authored in Bikeshed and published with embedded WebIDL blocks~\cite{gpuweb_spec}, which enables us to collect and normalize these definitions into a machine-readable API representation.
During parsing, the framework normalizes WebIDL constructs such as \texttt{interfaces}, \texttt{methods}, \texttt{typedefs}, \texttt{dictionaries}, \texttt{mixins}, and \texttt{enums} into a consistent schema used for rule extraction and mutation generation.

While WebIDL defines interface structure, it leaves certain semantic relationships underspecified. A noteworthy challenge is the use of flag-like namespaces, as shown in Listing~\ref{lst:gpu-buffer-usage}, in which bitfield constants are defined in separate \texttt{namespace} blocks but are intended to match \texttt{unsigned long typedefs}.
\begin{lstlisting}[style=code, language={C++}, caption={Example of flag-like namespace definitions.}, label={lst:gpu-buffer-usage}, captionpos=b]
typedef unsigned long GPUBufferUsageFlags;
namespace GPUBufferUsage {
  const GPUFlagsConstant COPY_SRC = 0x0004;
  // ... (other flags)
}
\end{lstlisting}

These associations are not explicit in the specification. \Sname therefore recovers them by associating \texttt{unsigned long typedefs} with their corresponding flag constants declared in specification-level namespaces. For example, constants such as \texttt{MAP\_READ} and \texttt{MAP\_WRITE} under \texttt{GPUBufferUsage} are merged into enumeration-like flag sets while preserving their numeric values. This normalization enables type-aware flag mutation.

\subsection{LLM-Assisted Implicit Rule Generation}
WebIDL captures only interface structures, and omits many constraints that appear only in the specification descriptions written in natural language. These include object lifecycles, command-ordering requirements, resource state transitions, and numeric alignment constraints (e.g., \texttt{bytesPerRow} must be a multiple of 256 in \texttt{GPUTexelCopyBufferInfo} validation~\cite{webgpu}). To recover these semantics, \Sname uses GPT-5 to identify candidate requirements in the WebGPU specification text. We then manually translate the outputs into rules that describe valid command sequences, required preconditions, and expected state transitions among WebGPU objects.

We model each WebGPU object (e.g., \texttt{GPUBuffer}, \texttt{GPUCommandEncoder}, \texttt{GPUQueue}, and \texttt{GPUCanvasContext}), as a state machine
whose state transitions are governed by API calls.
For example, a \texttt{GPUBuffer} may evolve through the lifecycle \texttt{unmapped} → \texttt{mapping} → \texttt{mapped} → \texttt{unmapped} → \texttt{destroyed}.
The inferred behavioral rules encode preconditions like the requirement that \texttt{writeBuffer()} can only be called when the buffer's usage flags include \texttt{COPY\_DST}, and postconditions such as
transitioning a command encoder to a finished state after calling \texttt{finish()}.

The rules also capture numeric layout and ordering constraints, such as parameter alignment requirements (e.g., \texttt{bufferOffset} must be a multiple of 4), as well as execution-order constraints (e.g., \texttt{setPipeline()} must precede any \texttt{draw*()} operation within a render pass).

All of these relationships are encoded symbolically using predicates such as \texttt{state\_eq} (state equality checks), \texttt{flags\_include} (bit-flag requirements), \texttt{order\_requires} (ordering dependencies), and \texttt{multiple\_of} (alignment constraints).
This representation separates candidate constraints from the mutation handlers. The prototype applies a supported subset to literal object arguments using field-name matching, without operation-aware resolution or interpretation of the complete state-machine model.
This implicit rule layer complements the WebIDL-based constraints and feeds directly into the AST mutation engine in the next stage.

\subsection{Rule-Guided AST Mutation}
After constructing the explicit and implicit rules, \Sname applies them to mutate the WebGPU CTS. The bundled CTS repository contains 3,411 tests (as of April 23, 2025), each designed to exercise a specific API behavior or validation rule.
We design an AST-guided mutator that programmatically modifies existing CTS tests according to selected extracted rules. Starting from CTS reuses the tests' established structure and harness integration, while mutation explores additional field and value configurations.

\begin{figure}[t]
    \centering
    \includegraphics[width=0.85\linewidth]{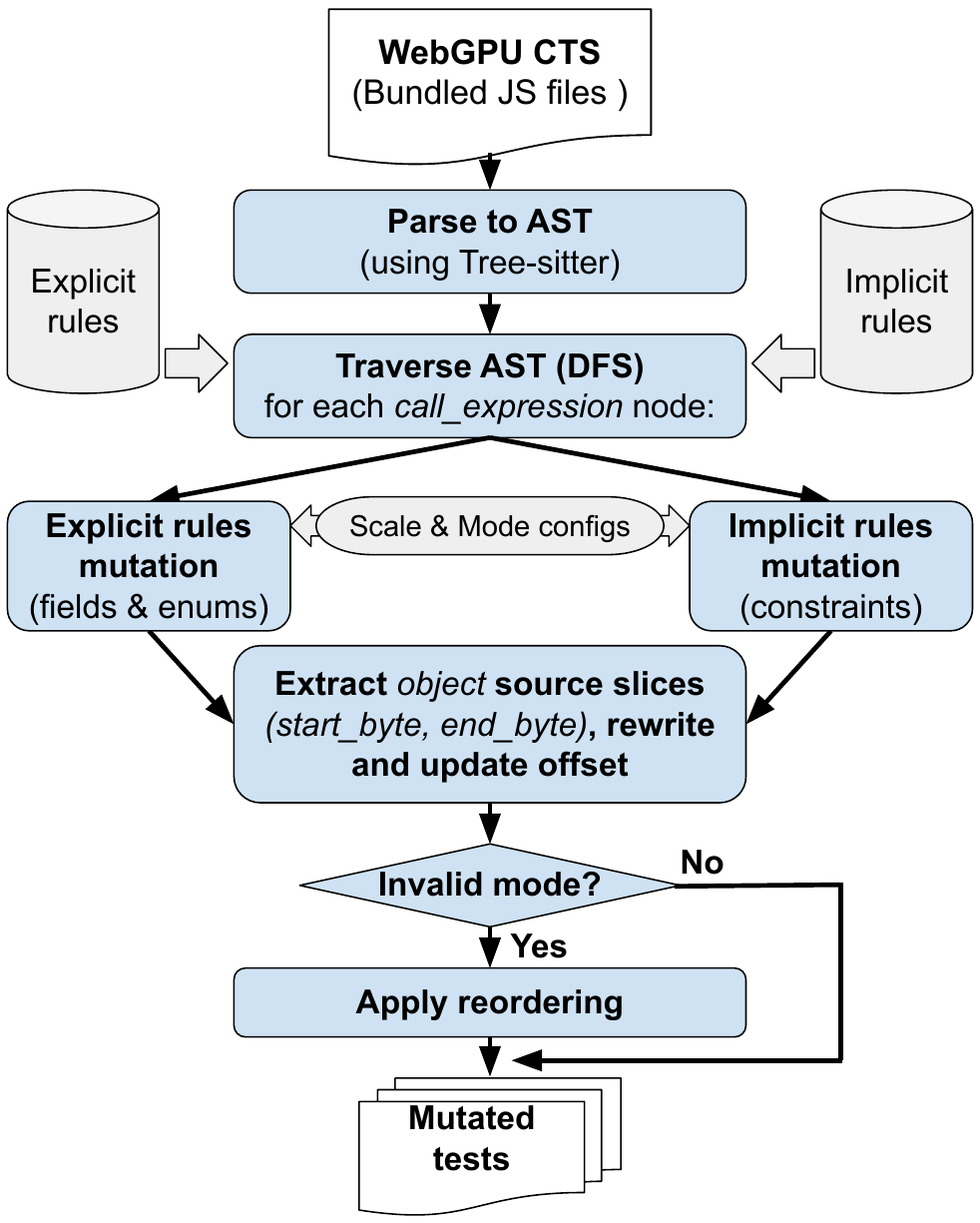}
   \caption{\Sname's AST-based mutation workflow.
}
    \label{fig:AST-Mutator}
\end{figure}

As shown in Figure~\ref{fig:AST-Mutator}, \Sname parses each CTS test file into an AST using Tree-sitter~\cite{tree-sitter}, and then applies a series of transformation passes.
Each pass corresponds to a type of rule application, such as inserting a required dictionary field, modifying an alignment-sensitive integer, or changing a textual flag combination. The transformations are probabilistically controlled by a mutation scale parameter, which sets the selection probability at each rule check, and a mode (valid or invalid) that selects the direction of handlers that implement mode-specific behavior.

More specifically, \Sname uses WebIDL-derived dictionary and
enumeration rules, together with selected implicit-rule handlers, to
guide mutations of object-literal arguments. For explicit rules, \Sname probabilistically inserts fields marked as
required when their names are absent from an object literal. This
insertion is performed in both modes. Its enumeration handler retains
a recognized enumeration string in valid mode and may substitute a
different recognized member in invalid mode.

For implicit rules, the implemented numeric handlers adjust integer
fields
according to constraints such as \texttt{multiple\_of} and \texttt{min\_value}, while the forbidden-pair handler modifies textual flag
combinations. These handlers use the field name of a rule's
target to locate candidate values within object literals.

The mutation loop parses each JavaScript input with Tree-sitter and visits \texttt{call\_expression} nodes. It examines direct object-literal arguments and applies the explicit rule set and supported implicit-rule handlers using textual field-name and value matching. Mutations are applied in traversal order. The scale is converted to a selection probability used at individual rule checks.

We note that our current ablation (Table~\ref{tab:ablation}) isolates the contribution of the two \emph{rule sources} (explicit vs.\ implicit) rather than the individual mutation handlers; disaggregating coverage and bug yield by handler is a natural extension we leave to future work, as it requires re-running the campaign with each handler disabled in turn.

\subsubsection{Configurable Validity Mode and Scale}
\label{sec:design:config}

\Sname generates static mutated suites under a chosen configuration and evaluates configurations offline using logs as the behavioral signal. This design avoids the code-coverage noise problem in the multi-layered and sandboxed browser graphics stack, and it also allows us
to produce static mutated suites that can be executed either by our executor or by the standard CTS harness on environments where real-time feedback is impractical, such as embedded Augmented Reality (AR) and Virtual Reality (VR) browsers. However, because WebGPU support on those platforms remains limited and these devices often run customized Chromium variants, we focus on testing Chromium.

The mode parameter reflects a broader hypothesis from prior dynamic analysis studies: programs may exhibit richer internal behavior when exercised with semantically valid inputs~\cite{aschermann2019nautilus}, or may reveal deeper flaws when intentionally stressed with invalid ones~\cite{oehlert2005violating}.

In valid mode, the supported numeric and forbidden-flag-pair handlers attempt to satisfy a selected local constraint; in invalid mode, they attempt to violate it. Required-field insertion is mode-independent, and the released rule encoding does not activate the prototype required-flag and ordering handlers. Thus, the mode names describe the intended direction of supported local mutations rather than certify an entire test as valid or invalid.
The scale parameter controls mutation intensity. Lower scales select fewer eligible edits in expectation; higher scales select more, which can increase diversity but also increase the likelihood of validation failures in shallow layers of the WebGPU implementation.

\subsection{Execution and Crash Analysis}

In the final stage, \Sname executes mutated tests on Chromium with WebGPU and runtime checks enabled (e.g., ASan and DCHECK). The executor serves mutated test cases via a lightweight local HTTP server and launches them across Chromium instances.
The executor orchestrates
browser lifecycles and collects each instance's console logs, including stack dumps and sanitizer prints.
This setup allows multiple tests to run in parallel, maximizing resource utilization during large campaigns.
Whenever a bug is found, our framework records a detailed crash artifact and saves the mutated test-case information needed for proof-of-concept (PoC) replay.

\section{Implementation}
\label{sec:impl}
We implemented \Sname as a framework that integrates specification extraction, rule-guided mutation, and test execution. The current prototype consists of approximately 1,300 lines of Python and shell scripts, excluding extracted artifacts and other data files.

\subsection{Explicit Rule Extractor}

\Sname extracts explicit syntactic and structural constraints directly from the W3C WebGPU specification dated August 5, 2025 as the ground-truth definition of
well-formed API usage. The
specification is written in the Bikeshed format.
The extraction procedure first builds the WebGPU specification locally using Bikeshed and then parses the generated HTML to collect all embedded WebIDL blocks. These fragments are cleaned, normalized, and merged into a single unified IDL file to eliminate duplication and resolve cross-references introduced by partial interfaces and mixins.

From the normalized IDL, \Sname extracts a structured representation of explicit API rules, including interface definitions, method signatures, argument types, dictionary fields, enumerations, typedefs, and mixin inclusion relationships. These rules are serialized into JSON and serve as the syntactic constraint layer for later mutation. Listing~\ref{lst:explicit-rules} shows a simplified example.

\begin{lstlisting}[
  language=json,
  caption={Example explicit rule representation.},
  label={lst:explicit-rules},
  float=t,
  captionpos=b
]
{
 "interfaces": {"GPUDevice": {
  "methods": {"createBuffer": {
   "args": [{"name": "descriptor",
             "type": "GPUBufferDescriptor"}]}}}},
 "dictionaries": {"GPUBufferDescriptor": {
  "size": {"type": "GPUSize64", "required": true},
  "usage": {"type": "GPUBufferUsageFlags", "required": true}}},
 "namespaces": {"GPUBufferUsage": {
  "kind": "flags",
  "constants": [{"name": "MAP_READ", "value": "0x0001"}]}}
}
\end{lstlisting}

The resulting rule set captures only constraints defined by the specification; semantic assumptions, object state transitions, and API ordering requirements are handled separately by the implicit rule layer.

\subsection{Implicit Rule Extractor}
\label{subsec:Implicit Rule Extractor}

For WebGPU usage constraints that appear only in the form of natural-language description,
\Sname uses an LLM (GPT-5) to assist in identifying candidate behavioral rules from selected excerpts of the WebGPU specification, including lifecycle requirements, state-dependent validity conditions, and explicit ``must'' or ``must not'' statements.
We prompt the model to translate each requirement into a
testable rule by specifying the affected API object(s), the relevant precondition(s), and the required or forbidden operation(s). The LLM is used only for the extraction of implicit rules from the specification and is not otherwise involved in generating, mutating, or executing test cases.

The model outputs are manually post-processed to remove duplicates, normalize terminology to match WebIDL identifiers, and discard vague or non-actionable statements. The remaining candidates are manually reviewed against the specification and then encoded in JSON for the mutation engine. We estimate that this process takes a developer proficient in the WebGPU API less than 20 hours. Listing~\ref{lst:implicit-rules} shows a simplified example.

\begin{lstlisting}[
language=json,
caption={Example implicit rule representation.},
label={lst:implicit-rules},
captionpos=b
]
{
  "state_machine": {"GPUBuffer": {
    "states": ["unmapped", "mapping", "mapped", "destroyed"],
    "transitions": [
      { "on": "mapAsync(start)", "from": "unmapped", "to": "mapping" },
      { "on": "unmap", "from": "mapped", "to": "unmapped" }]}},
  "ops": [
    {
      "op": "GPUBuffer.mapAsync",
      "requires": [
        { "kind": "state_eq", "target": "this.state", "equals": "unmapped" }
      ],
      "effects": [
        { "kind": "post_state", "target": "this.state", "to": "mapping" }
      ]
    }
  ]
}
\end{lstlisting}

\subsection{AST-Level Mutation Engine}

\Sname mutates WebGPU CTS programs by using a JavaScript AST to locate candidate source regions. We first build the TypeScript CTS source into JavaScript and then use Tree-sitter~\cite{tree-sitter}'s JavaScript parser to locate call expressions and their direct object-literal arguments. We do not reparse outputs to reject syntax errors because they may be valuable fuzzing inputs.

The mutator takes as input (1) a CTS JavaScript test file, (2) explicit WebIDL-derived rules, (3) implicit semantic and state rules encoded in JSON, and (4) the mode and scale configurations. It traverses the AST to locate direct object-literal arguments of call expressions and applies rules through textual field-name and value matching. Rule-guided edits insert fields marked as required, substitute recognized enumeration strings, adjust numeric parameters, or modify forbidden flag combinations.

\subsection{Test Execution and Reproducibility}

Although the mutated CTS can already run within the CTS harness, we implement a test execution framework for automatic dynamic testing. \Sname runs mutated CTS suites on an ASan- and DCHECK-instrumented Chromium build with WebGPU enabled.
Each mutated test directory is served on a local web server. The executor launches a Chromium process per CTS query and navigates to the local runner URL. Each invocation has a 15-second timeout.

\input{eval}

\section{Related Work}

\subsection{Web-Based GPU Testing and Security Analysis}
The security and reliability of web-exposed GPU interfaces have been examined from multiple angles, ranging from web app isolation to microarchitectural side channels.
Rendered Insecure~\cite{naghibijouybari2018rendered} demonstrated timing- and cache-based side channels through WebGL, while more recent work such as Drive-by GPU~\cite{giner2024generic}, WebGPU-SPY~\cite{ferguson2024webgpu}, and AtomicIncrement~\cite{hohentanner2025unveiling} showed that similar channels could be exploited remotely through WebGPU compute shaders.
Sugar~\cite{yao2018sugar}, Milkomeda~\cite{yao2018milkomeda}, and Belkin et al.~\cite{belkin2019risks}
studied WebGL isolation and security checks; Milkomeda specifically repurposes
browser-enforced WebGL checks to protect mobile GPU interfaces.
LockedApart~\cite{laor2025poster} reveals how thread contention can fingerprint GPU microarchitectures through WebGPU.
These studies highlight the dual nature of web graphics APIs’ programmability, offering powerful client-side computation while also broadening the attack surface that requires systematic and automated testing. This is precisely the problem that \Sname addresses.

TwinFuzz~\cite{leonelli2025twinfuzz} tests video hardware acceleration using differential outputs and coverage from a white-box software decoder to guide testing of hardware decoding.

\subsection{Fuzzing Frameworks for Web Graphics}
Fuzzing has been widely used to discover bugs in web interfaces, although traditional coverage-guided fuzzers often struggle with web graphics APIs (\S\ref{sec:intro}). Prior work has primarily targeted WebGL or shader compilers.
GLeeFuzz~\cite{peng2023gleefuzz} guides mutations using runtime error messages to reach deeper WebGL rendering paths.
WGSLSmith~\cite{mohsin2022wgslsmith} generates random WebGPU shaders (in WebGPU Shader Language, WGSL) to test shader compiler paths, while DarthShader~\cite{bernhard2024darthshader} targets the WGSL compilation pipeline using a hybrid approach combining intermediate representation (IR) and AST-based mutations.
Since WGSL is a separate shader language and is treated as data rather than API-level code in WebGPU, \Sname focuses on API-level semantics, which is orthogonal to shader testing.

Beyond the graphic stack, general browser fuzzers have also targeted WebIDL-defined interfaces. Domino~\cite{kratzer2020fuzzing} uses Firefox's own WebIDL bindings as a fuzzing grammar to generate JavaScript code. Minerva~\cite{zhou2022minerva} improves on prior generation-from-scratch browser fuzzers by mining dynamic modify-reference relations between API calls. Favocado~\cite{dinh2021favocado} targets the binding layer between JavaScript engines and native code, generating semantically correct test cases by extracting type and state information for binding objects. These systems share \Sname's goal of respecting API-level semantics to reach deeper implementation code, but \Sname mutates existing CTS tests rather than generating programs from scratch.

WG-Fuzz~\cite{simols2024_wg_fuzz} is a differential fuzzer for WebGPU JavaScript APIs that generates API call sequences guided by a resource-aware state model and reports 6 crashing bugs.
WebGlitch~\cite{wong2025webglitch} generates valid-by-construction WebGPU programs using a curated precondition model.

We compare the approaches along three axes. First, \textbf{generation strategy}: rather than synthesizing programs from an API model, \Sname starts from the existing CTS corpus and mutates it, reusing established API setup and call sequences; the Skia case in \S\ref{sec:motivation} illustrates the complex browser behavior reachable from these tests. Second, \textbf{rule acquisition}: WebGlitch's correctness model is built entirely by hand and requires continual manual maintenance as the specification evolves, whereas \Sname derives explicit rules automatically from WebIDL and uses LLM assistance to propose implicit semantic rules from specification prose, which are then manually reviewed and encoded for the mutator. Third, \textbf{mode comparison}: WebGlitch treats invalid-call injection as a secondary, probabilistic option layered on top of its primarily valid-by-construction generation, while \Sname compares valid and invalid mutation modes across scales and reports their observed log-location counts and bug yield (\S\ref{sec:evaluation}).

The two approaches are complementary rather than competing: WebGlitch's from-scratch, on-demand generation explores a broad space of self-contained programs across two independent WebGPU implementations and their JavaScript-runtime bindings, while \Sname's mutation-based approach targets deep, state-dependent paths already reachable by CTS but left untested by conformance checks alone, within Chromium. \Sname discovers 7 distinct issues in Chromium and, in at least one case, exercises a browser call chain exceeding 100 stack frames; we use this observation as qualitative evidence that CTS-anchored mutation can expose complex browser behavior. Combining WebGlitch-style from-scratch generation with \Sname-style specification-grounded mutation is a promising direction for future work.

\subsection{Type-Safe WebGPU Implementations}

The wgpu project~\cite{wgpu_rust} uses Rust’s type system and ownership model to improve safety in WebGPU backends. This does not eliminate the need for correct constraint enforcement in browser-facing C++ components and privileged GPU-process code, so \Sname remains complementary even in partly type-safe implementations.

\subsection{AI for Specification Understanding}
Recent advances in AI-guided fuzzing demonstrate the potential of LLMs for understanding structured specifications and inferring program semantics~\cite{rahman2024cellularlint, yao2024formal, li2025extracting, nam2024using, nikbakht2024tspec}.
MetaMut~\cite{ou2024mutators} uses LLM-generated mutation operators for compiler fuzzing, achieving greater coverage without human-written templates.
HLPFuzz~\cite{yang2025hybrid} uses LLM-based constraint solving, hybrid centrality prioritization, and iterative context construction to reach deeper program states in language processors.
ChatAFL~\cite{meng2024large} uses pretrained LLM knowledge to construct protocol-message grammars and predict messages.
Failure Modes of Standalone LLM-Based Test Generation~\cite{yao2026failure}
argues that LLM-generated tests can encode faulty behavior when accepted without
independent oracles or human review.  \Sname avoids standalone LLM-based test generation through manual inspection against the specification.
These works demonstrate the feasibility of using AI for specification understanding and related software-engineering tasks. \Sname
derives explicit syntactic rules directly from the official WebIDL and augments them with implicit behavioral constraints inferred from the specification text.
This combination grounds LLM-assisted specification extraction in a syntactic layer suitable for test generation.

\section{Discussion and Future Work}
\label{sec:discussion}

\Sname demonstrates that CTS-anchored, specification-guided mutation is a practical way to exercise deep, state-dependent behaviors and to uncover failures not triggered by the unmodified CTS. We view this approach as a foundation for dynamic conformance testing that tracks an evolving standard.

\begingroup
\textbf{Limitations.} We highlight five limitations of the current study. First, our primary behavioral-diversity metric---unique log-emitting locations---is a partial proxy for execution diversity rather than a substitute for instrumented code coverage (\S\ref{subsec:setup}); the graphics-related subset of these locations is also small and grows only modestly with mutation scale (\S\ref{subsec:singleanalysis}), so claims about deep graphics-stack exploration should be read as suggestive rather than exhaustive. Second, our ablation (\S\ref{sec:ablation}) isolates the two rule \emph{sources} (explicit vs.\ implicit) rather than individual mutation handlers; it cannot attribute coverage or bug yield to a particular transformation. Third, four of the seven issues in Table~\ref{tab:issues} did not reproduce
possibly due to nondeterministic browser behavior.
Fourth, the implicit rules that drive semantics-aware mutation are extracted with GPT-5 assistance and then manually validated by the authors (\S\ref{subsec:Implicit Rule Extractor}); this manual validation step introduces subjectivity that an independent audit could help quantify. Fifth, our evaluation targets a single browser engine (Chromium) and its WebGPU implementation (Dawn); we have not evaluated portability to Firefox, Safari, or the wgpu-based implementations that other WebGPU fuzzers have tested, so cross-implementation generalization remains an open question.
\endgroup

\textbf{Future directions.} \Sname currently targets WebGPU, but the core workflow of extracting constraints from a written standard and using them to mutate a conformance suite is not WebGPU-specific.
\Sname leverages an LLM to propose candidate semantic rules and then manually validates them. We expect the manual effort to shrink as models improve and as we add automated checks that tie each rule to a specific specification requirement and detect when specification updates invalidate a rule.
A future direction is an end-to-end pipeline that can ingest a new web API specification and its conformance suite and produce a mutation-based tester with little or no human curation.

\textbf{Implications for the community.} Our findings reinforce that
conformance suites are necessary but not sufficient for complex, stateful APIs. Even when all original CTS tests pass, specification-derived mutations can expose additional memory-safety bugs. Treating the specification as a source of executable constraints makes it possible to explore long call chains into deeper program states. We expect dynamic, specification-driven mutation to complement both manual CTS development and WebGPU fuzzers.

\section{Conclusion}
We present \Sname, a specification-guided dynamic conformance testing framework for WebGPU built on the official CTS. \Sname extracts syntactic API constraints from WebIDL in the specification, derives semantic rules with LLM assistance, and uses selected rules to guide AST-located textual mutations under two intended-direction modes. Running these suites on ASan-instrumented Chromium~133 and~146 triggered seven issues, three of which we reproduced and responsibly reported to Google, including a heap corruption in V8's thread-isolation allocator and a security assertion failure that the vendor confirmed as duplicated internally-known bugs that were unpatched at the time of our findings. These results show that specification-grounded mutation can extend CTS beyond fixed patterns and reach deeper, cross-component behaviors, even when---as here---some of those behaviors originate outside the WebGPU implementation itself (e.g., in V8, Mojo, or Skia).

\section*{Acknowledgments}
\label{sec:acknowledgments}
We acknowledge the use of ChatGPT 5 to extract the implicit rules described in Section~\ref{subsec:Implicit Rule Extractor}, and to improve the grammar and readability of this manuscript.
This work received no external
funding.

\bibliographystyle{IEEEtran}
\bibliography{references}

\end{document}

%% file: eval.tex
\section{Evaluation}
\label{sec:evaluation}

Our evaluation answers four questions: (1) How do mutation \emph{mode} (valid vs.\ invalid) and \emph{scale} affect the diversity of WebGPU behavior exercised by \Sname? (2) How quickly does the fuzzer discover new log-emitting locations, and which scales provide the best trade-off between exploration depth and efficiency? (3) What is the impact of running multiple fuzzers in parallel, and does it increase or decrease behavioral coverage? (4) How does Chromium~146 compare with Chromium~133 on the same mutated suites? Unless otherwise noted, all experiments were conducted on Chromium~146 with ASan, DCHECK, and WebGPU enabled.
We treat DCHECK crashes as bug signals requiring further analysis because the checks are removed in production builds, allowing guarded conditions to proceed without the assertion.

\begin{figure*}[t]
    \centering
    \begin{minipage}[t]{0.45\textwidth}
        \centering
        \includegraphics[width=\linewidth]{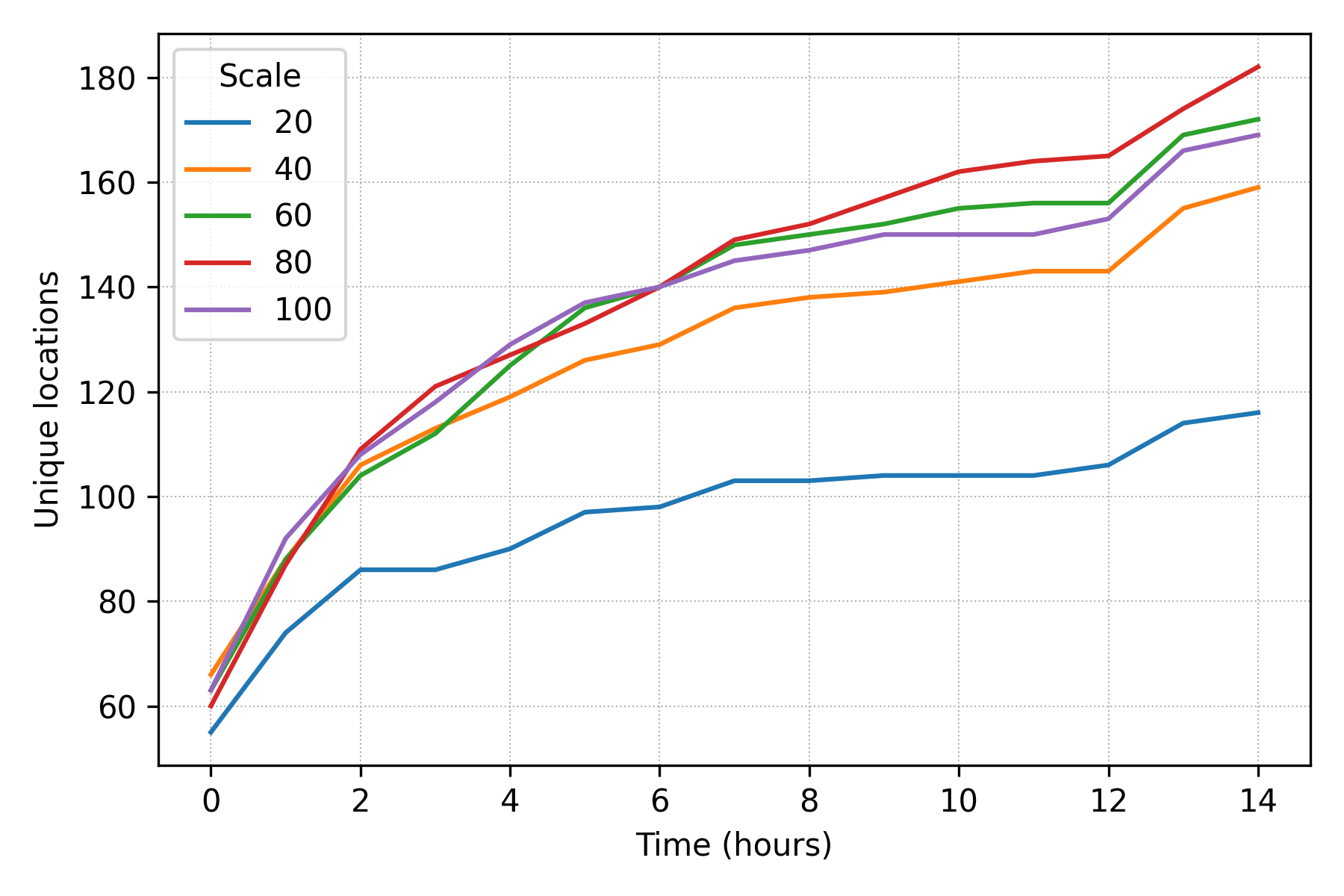}
        \caption{Valid-mode cumulative unique locations.}
        \label{fig:valid-coverage}
    \end{minipage}
    \hfill
    \begin{minipage}[t]{0.45\textwidth}
        \centering
        \includegraphics[width=\linewidth]{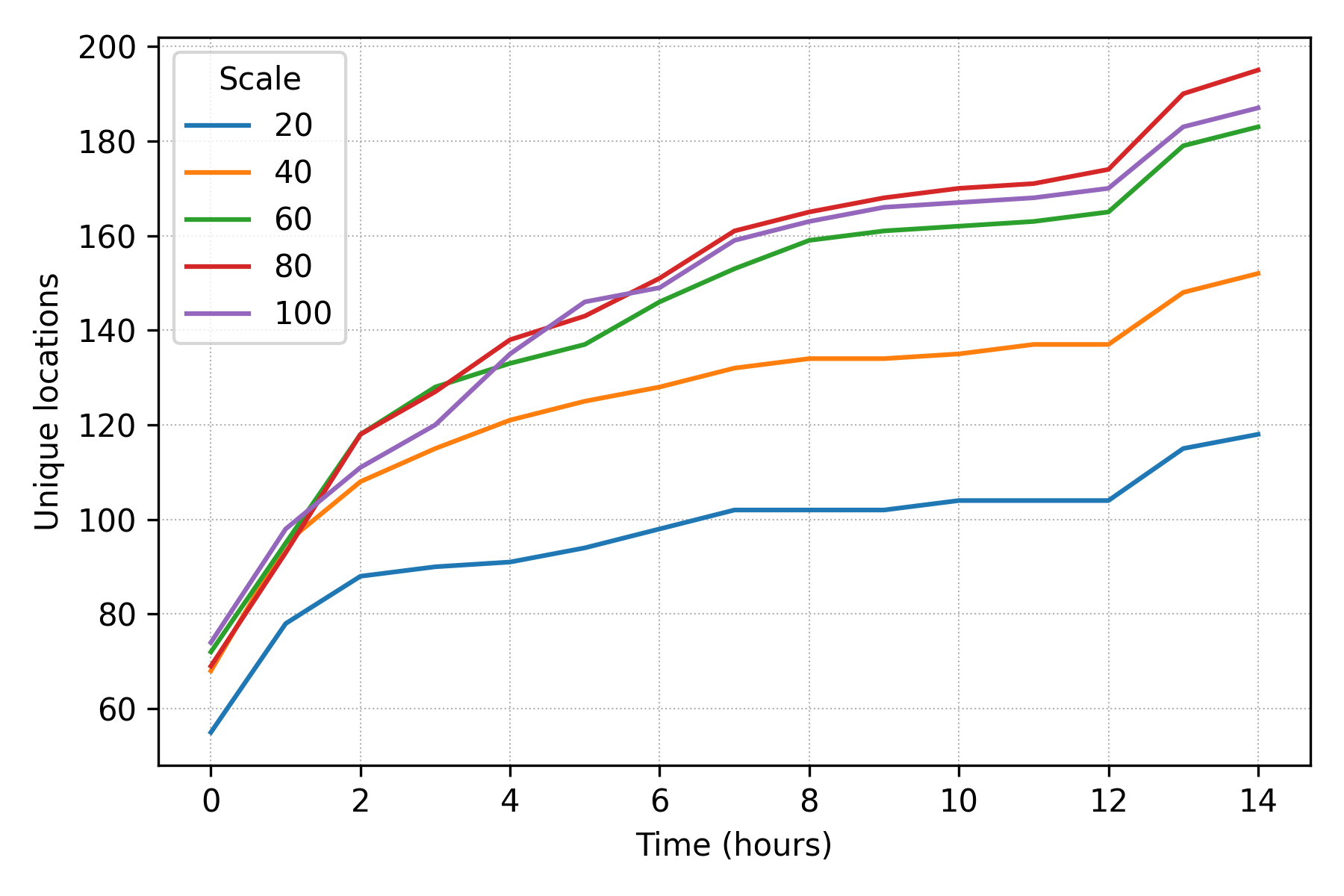}
        \caption{Invalid-mode cumulative unique locations.}
        \label{fig:invalid-coverage}
    \end{minipage}
    \vspace{0.3em}
    \begin{minipage}[t]{0.45\textwidth}
        \centering
        \includegraphics[width=\linewidth]{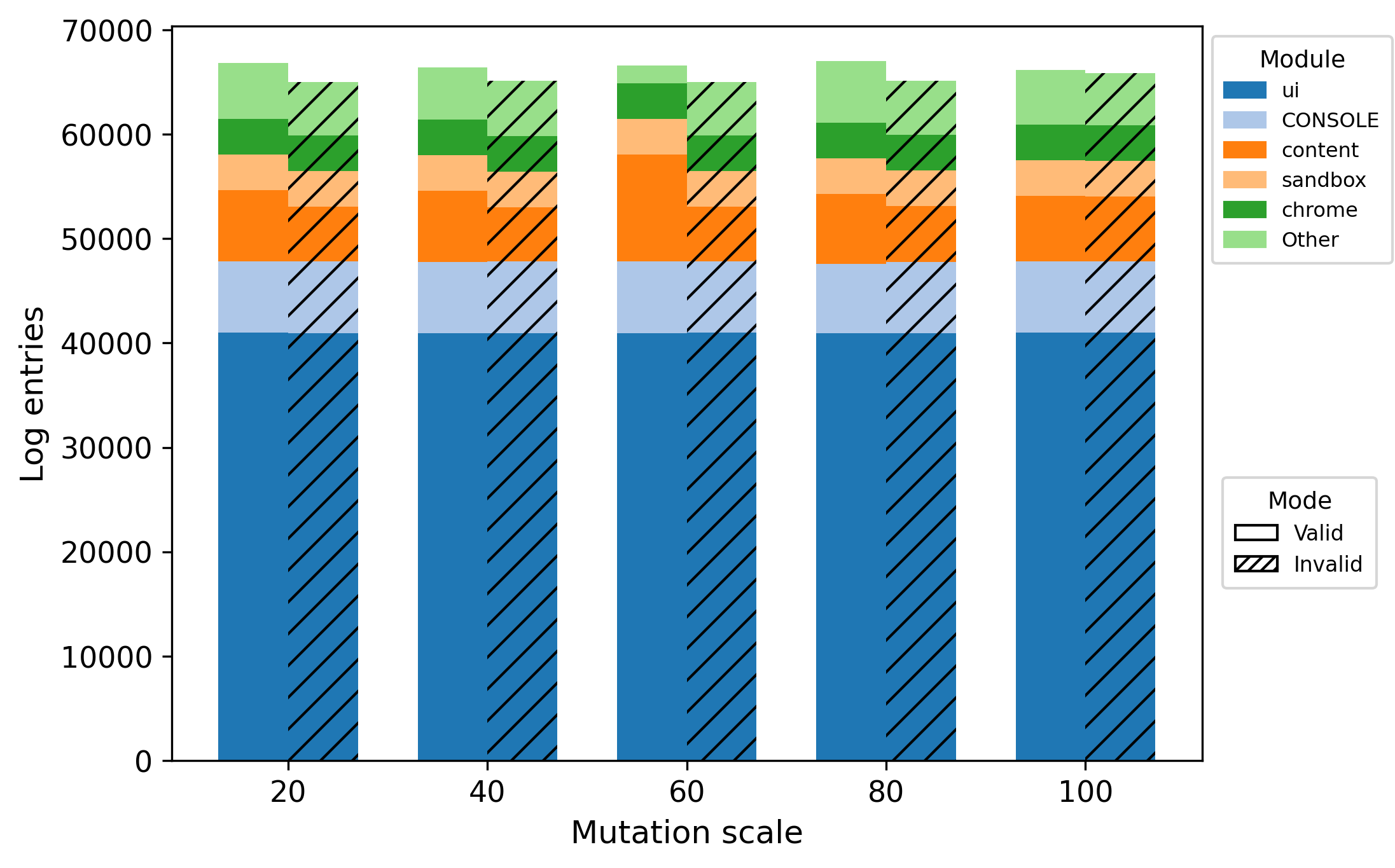}
        \caption{Cumulative log entries by module and mutation scale for valid and invalid modes.}
        \label{fig:module-breakdown}
    \end{minipage}
    \hfill
    \begin{minipage}[t]{0.45\textwidth}
        \centering
        \includegraphics[width=\linewidth]{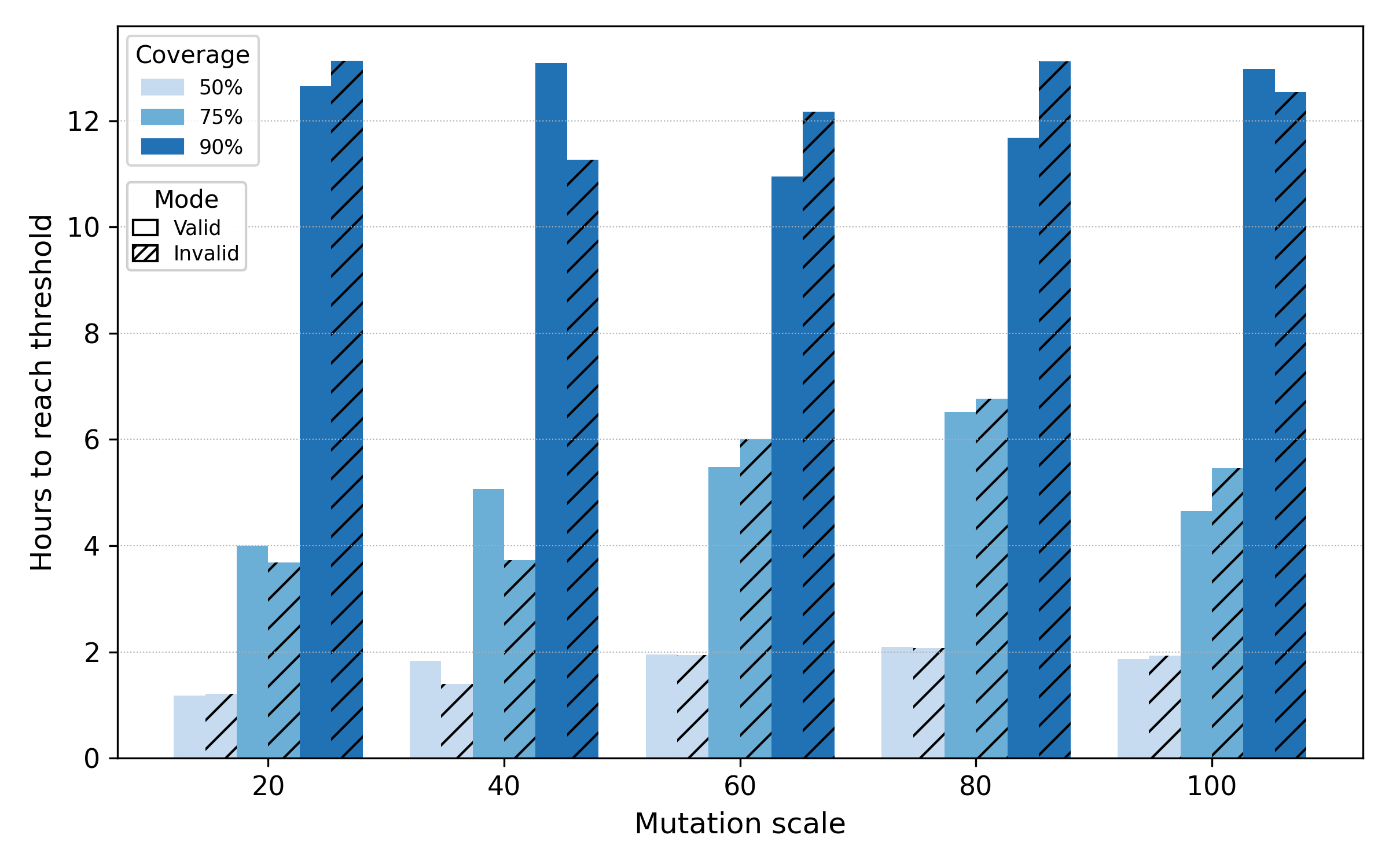}
        \caption{Hours to reach 50\%, 75\%, and 90\% of final unique locations for valid and invalid modes across scales.}
        \label{fig:time-milestones}
    \end{minipage}
\end{figure*}

\begin{figure*}[t]
    \centering
    \begin{minipage}[t]{0.45\textwidth}
        \centering
        \includegraphics[width=\linewidth]{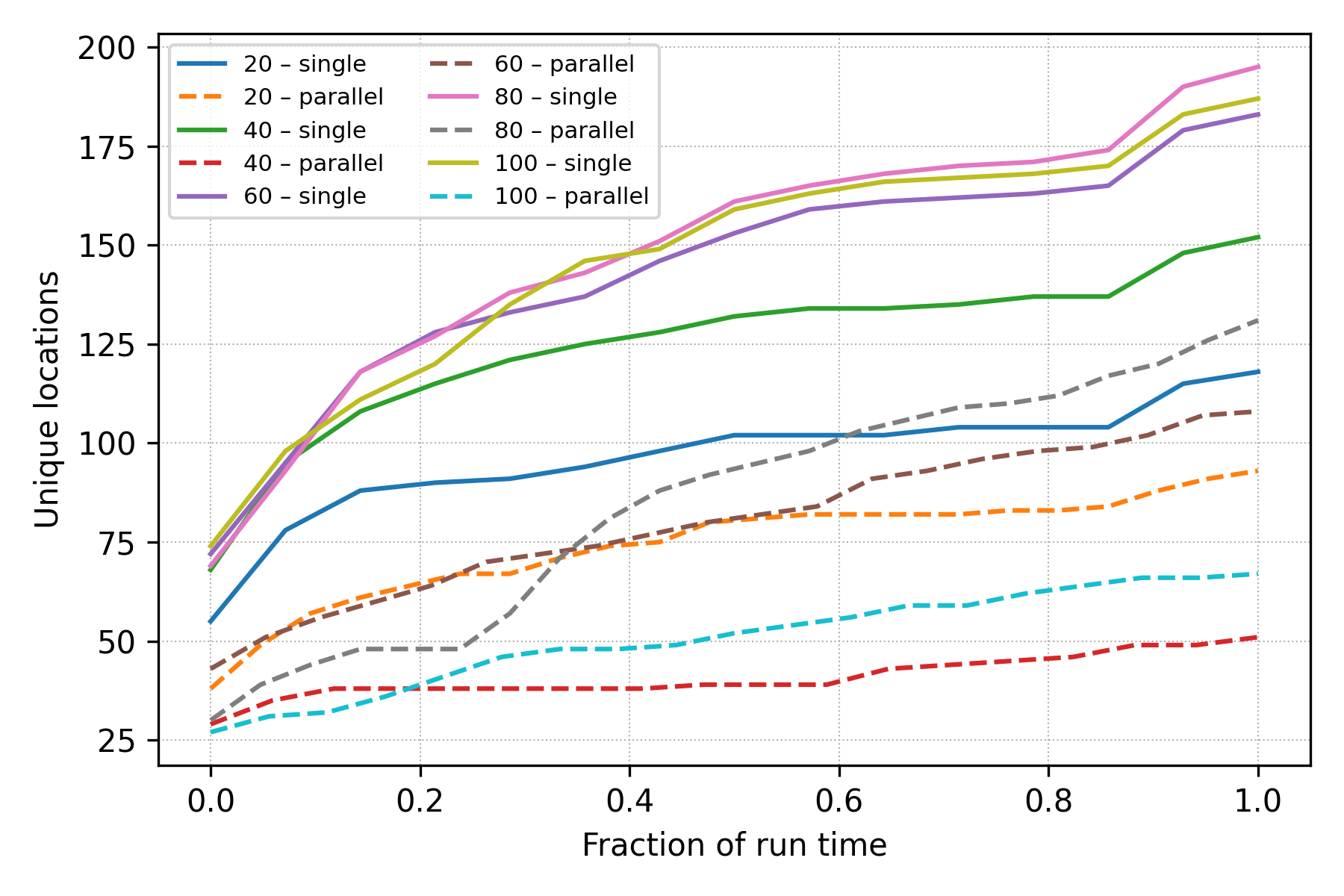}
        \caption{Single vs. parallel cumulative unique locations in invalid mode. Solid lines are single runs; dashed lines are parallel runs.}
        \label{fig:parallel-unique}
    \end{minipage}
    \hfill
    \begin{minipage}[t]{0.45\textwidth}
        \centering
        \includegraphics[width=\linewidth]{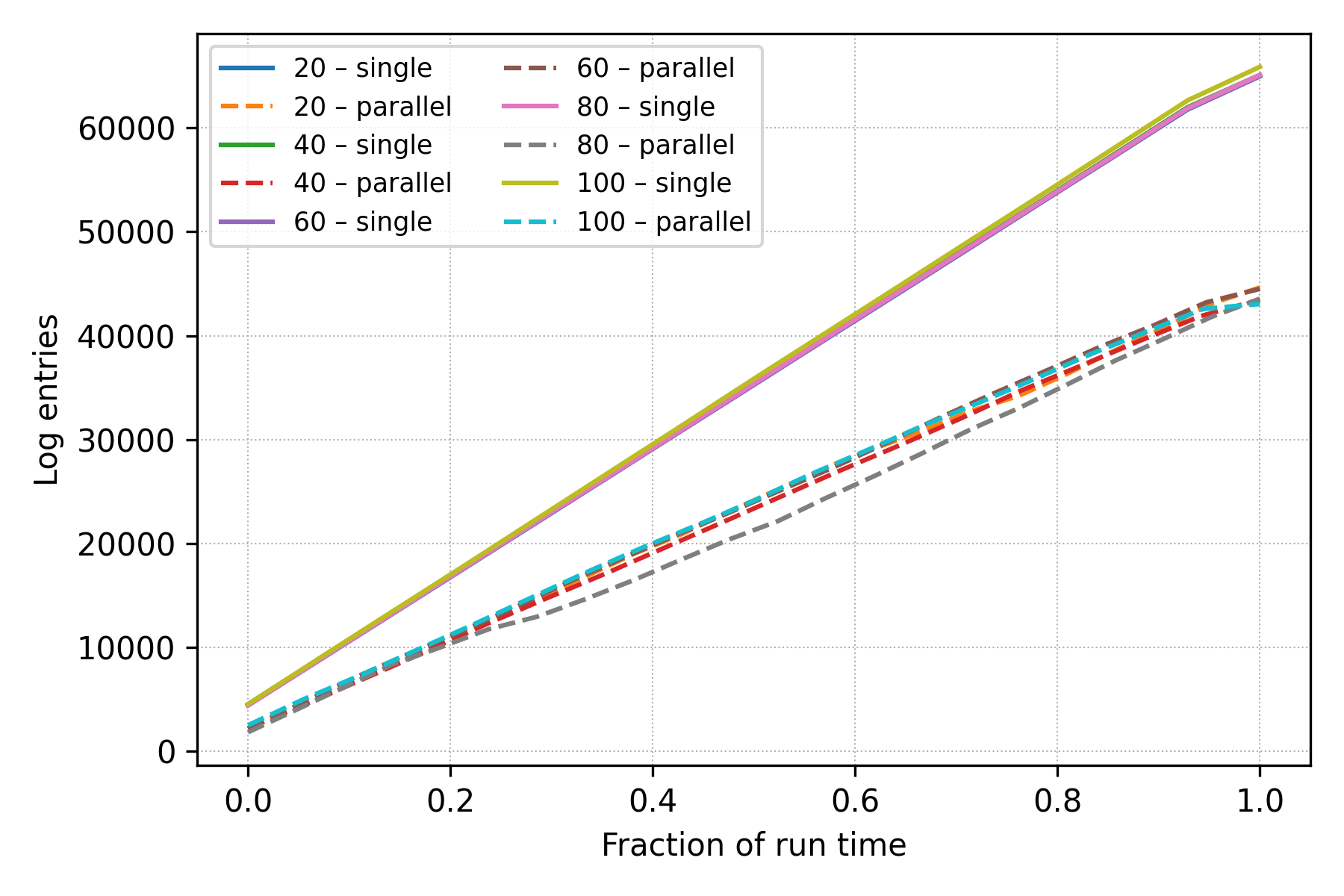}
        \caption{Single vs.\ parallel cumulative log entries in invalid mode. Solid lines are single runs; dashed lines are parallel runs.}
        \label{fig:parallel-logs}
    \end{minipage}
    \vspace{0.3em}
    \begin{minipage}[t]{0.45\textwidth}
        \centering
        \includegraphics[width=\linewidth]{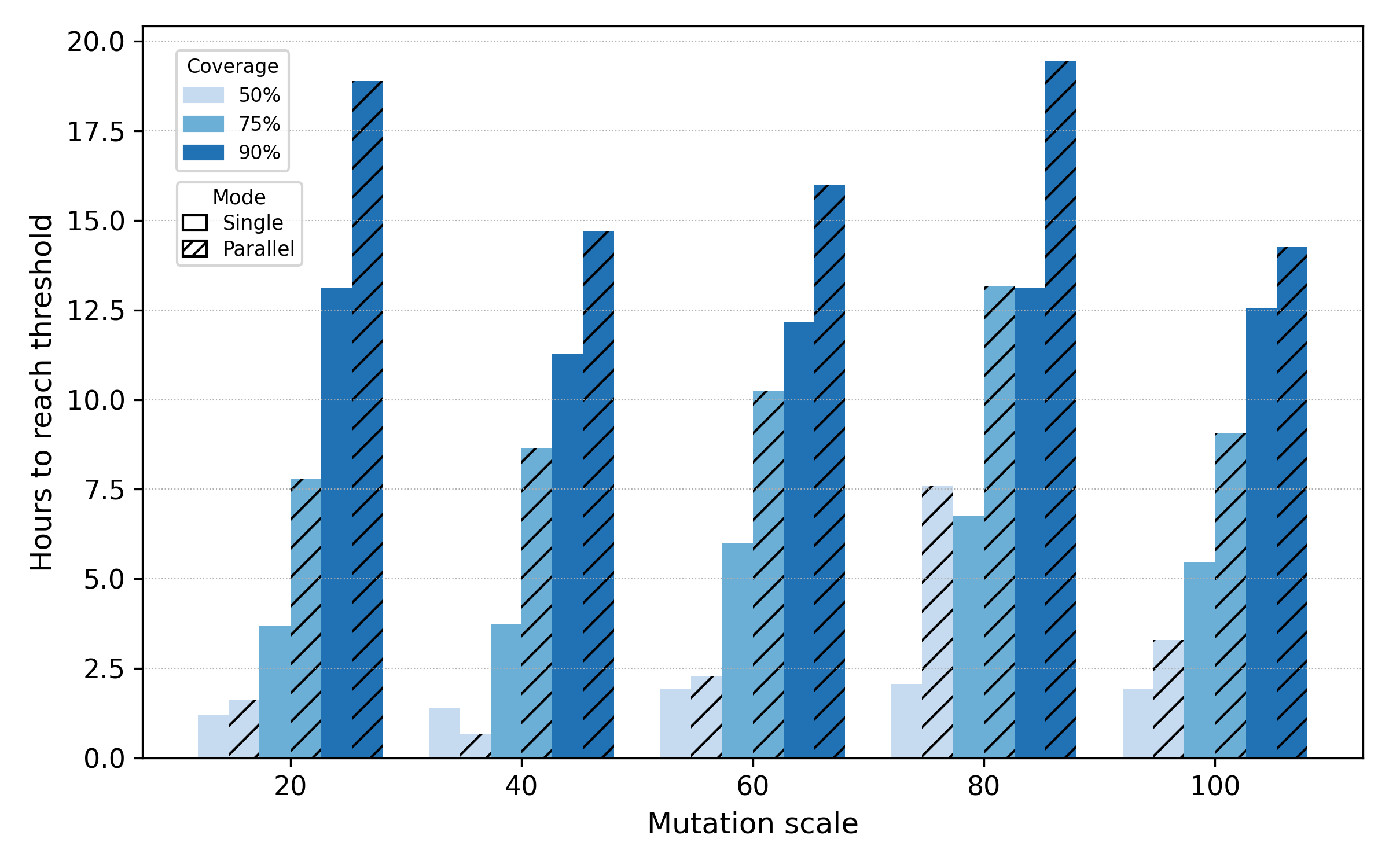}
        \caption{Time to reach coverage milestones for single and parallel runs in invalid mode. Bars indicate hours to reach 50\%, 75\% and 90\% of the final unique count.}
        \label{fig:parallel-ttc}
    \end{minipage}
    \hfill
    \begin{minipage}[t]{0.45\textwidth}
        \centering
        \includegraphics[width=\linewidth]{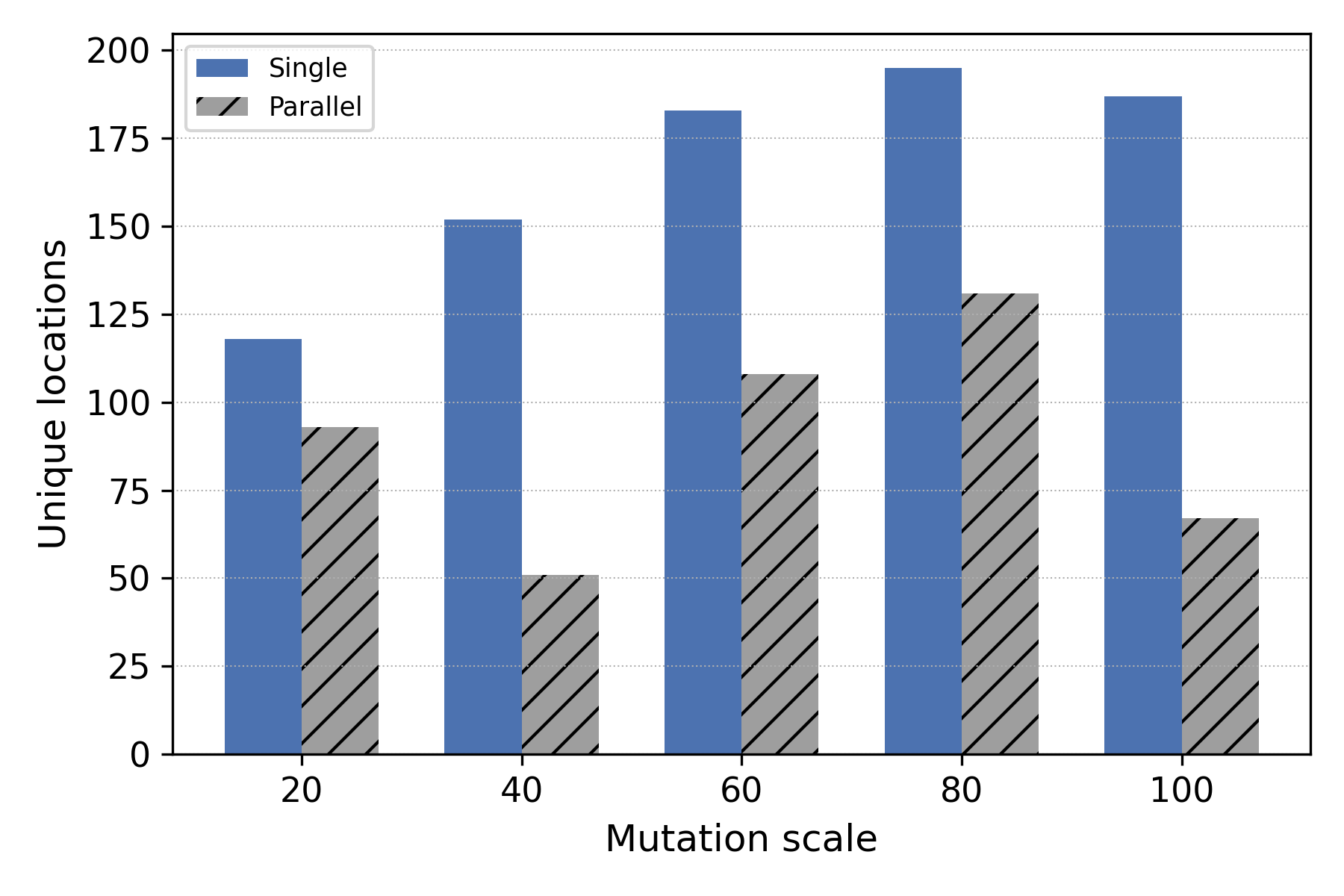}
        \caption{Final unique locations for single and parallel runs in invalid mode.}
        \label{fig:parallel-final}
    \end{minipage}
\end{figure*}

\subsection{Experimental Setup}
\label{subsec:setup}

As discussed in \S\ref{sec:design}, \Sname uses the CTS as its seed corpus and mutates existing tests.
For each scale in \{20, 40, 60, 80, 100\}, we generated two mutated suites: a \emph{valid} suite that preserves extracted rules and an \emph{invalid} suite that intentionally violates selected rules. A single run attempts the bundled CTS queries sequentially (no parallelism) and takes approximately 15 hours, including a 15-second timeout per query.
Each bracketed log line (e.g., \texttt{[pid:tid:timestamp:SEVERITY:file:line]}) yields a \emph{location key} (\textit{file.cc}:\textit{line}) and a severity level (INFO, WARNING, ERROR, FATAL). We treat each distinct location key as a unique log-emitting site and count subsequent occurrences as revisits.

Conventional code-coverage collection is difficult in our end-to-end setting because a WebGPU test executes across multiple isolated Chromium processes and graphics-stack components, requiring coverage information to be collected and combined across process boundaries---the same obstacle noted by prior work on runtime-signal-guided WebGL fuzzing~\cite{peng2023gleefuzz}.
We use unique log-emitting locations as a proxy for execution diversity, similar to prior work that uses runtime signals such as error messages to guide exploration in WebGL fuzzing~\cite{peng2023gleefuzz}.
Parallel runs execute six tests concurrently on the same machine using different mutated suites; we aggregate all log lines and sort them by their internal timestamp.

\subsection{Coverage Analysis}
\label{subsec:singleanalysis}

\textbf{Coverage growth.} Figures~\ref{fig:valid-coverage} and \ref{fig:invalid-coverage} plot the cumulative number of unique log-emitting locations for valid and invalid modes, respectively.
\begingroup
The x-axis reports absolute execution time in hours, allowing the coverage growth
and testing cost of each configuration to be observed directly.
\endgroup
Each curve is monotonic because coverage is cumulative, and both modes continue to discover new locations throughout the run.
Across the scales we tested, invalid mode generally reaches a higher final count of unique log-emitting locations than valid mode.
At scale~80, where both modes reach their peak number of unique locations, valid mode reaches 182 unique locations, whereas invalid mode reaches 195.
This indicates that controlled semantic violations exercise a broader set of validation and program states.
Increasing the mutation scale beyond~80 yields diminishing returns.

\textbf{Time-to-coverage milestones.} To quantify discovery speed, we measured the time (in hours) to reach 50\%, 75\% and 90\% of the final unique count for each mode and scale.
Figure~\ref{fig:time-milestones} shows that milestone times are broadly similar across configurations, though scale-specific differences remain visible.
For example, at scale~60, the valid run needs 10.9~h to reach 90\% coverage, compared to 12.2~h for the invalid run.
At scale~100, the valid run reaches 90\% coverage in about 13.0~h, while the invalid run reaches it in about 12.5~h. At lower thresholds, the same variability remains visible: the 50\% milestone is reached in roughly 1.2--2.1 hours across all settings, while the 75\% milestone ranges from about 3.6~h to 6.8~h depending on scale and mode. Overall, the figure suggests that mutation scale affects not only final coverage but also the rate at which coverage accumulates, and that the most time-efficient configuration depends on the chosen coverage target. In particular, scale~60 valid mode appears especially efficient for reaching 90\% coverage.

\textbf{Module-level breakdown.} To understand the browser components explored by the tests, we classified each log line by the top-level directory of its C++ source path (e.g., \texttt{ui}, \texttt{chrome}, \texttt{sandbox}).
Note that \texttt{chrome} represents the core browser code responsible for the browser user interface and features such as bookmarks.
Figure~\ref{fig:module-breakdown} shows the cumulative log entries by module for valid and invalid modes across scales.
Across all configurations, the dominant contribution comes from the \texttt{ui} module, followed by \texttt{CONSOLE} and \texttt{content}, while \texttt{sandbox}, \texttt{chrome}, and \texttt{Other} contribute smaller but consistent portions.
The overall distribution of log entries remains stable across both mutation modes and scales. Variations across scales are minor, with scales~80 and~60 showing slightly higher total log counts, while scale~100 does not produce a noticeable increase despite its higher mutation intensity.

To better understand how \Sname exercises the browser graphics stack, we tracked unique log-emitting locations from graphics-related components. Across single runs, the number of graphics-related locations is small compared to the total number of log-emitting locations. In valid mode, scales~20, 40, and~60 each reach 15 locations, scale~80 increases slightly to 17, and scale~100 decreases to 14. In invalid mode, the count stays between 14 and 15 across all scales. These results show that increasing the mutation scale significantly increases overall coverage but has little effect on graphics-related coverage. Scale~80 in valid mode achieves the highest count, while invalid mode shows almost no variation across scales. This suggests that the graphics stack is exercised through a fairly fixed set of paths, with little expansion at higher mutation levels.

\begingroup
\subsection{Ablation Study}
\label{sec:ablation}

To quantify the contribution of the two specification-derived rule
layers, we perform an ablation study on Chromium~146 using the
single-run invalid configuration at mutation scale~80. We compare the
unmodified WebGPU CTS with three mutation configurations:
\emph{explicit-only}, which uses the WebIDL-derived structural rules;
\emph{implicit-only}, which uses the semantic constraints recovered
from the specification text; and the full \Sname configuration, which combines
both rule layers. We use the number of unique log-emitting source
locations as our primary measure of behavioral diversity and also
report the subset originating from graphics-related components.

Table~\ref{tab:ablation} shows an increase in observed log-location counts
when either rule layer is enabled. The unmodified CTS reaches 55 unique
log-emitting locations, whereas the implicit-only configuration reaches
82, an increase of approximately 49\%. The explicit-only configuration
reaches 187 unique locations, more than three times the CTS baseline.
This difference is consistent with the roles of the two rule layers:
the WebIDL-derived rules apply broadly to API descriptors, fields, and
enumerated values, and therefore expand the structural input space
substantially, while the implicit rules target a more selective set of
semantic relationships such as flag combinations, numeric constraints,
object states, and API ordering.

Combining the two layers yields the highest observed log-location count, with
195 unique locations. In particular, Full \Sname reaches locations
beyond those observed with the explicit layer alone, suggesting that
the semantic constraints can contribute additional exploration in this
comparison. Overall, the ablation
suggests that the explicit layer provides most of the breadth in the
explored API surface, while the implicit layer contributes additional
semantics-aware behaviors; their combination provides the broadest
observed logging footprint.

The graphics-related counts remain comparatively stable across the four
configurations (15--19 locations), consistent with the trend observed in
our scale analysis. There, the number of graphics-related locations
changes only slightly across mutation scales, even when the total number
of unique log-emitting locations increases substantially. This suggests
that the additional behavioral diversity introduced by the
specification-derived rule layers extends beyond graphics-specific
logging sites to other browser components exercised during WebGPU
CTS workloads.
Our case studies (\S\ref{subsec:asan-case}) likewise illustrate bugs in other browser components during mutated CTS execution.

\begin{table}[t]
\centering
\caption{Ablation of \Sname's specification-derived rule layers on
Chromium~146 using a single invalid-mode run at mutation scale~80.}
\label{tab:ablation}
\begin{tabular}{lrr}
\hline
\textbf{Configuration} &
\textbf{Unique Locations} &
\textbf{Graphics-Related} \\
\hline
Original CTS        & 55  & 18 \\
Implicit rules only & 82  & \textbf{19} \\
Explicit rules only & 187 & 15 \\
Full \Sname        & \textbf{195} & 15 \\
\hline
\end{tabular}
\end{table}

This ablation isolates the contribution of the two \emph{rule sources} (explicit WebIDL constraints vs.\ implicit specification-prose constraints) rather than individual mutation handlers. Each rule source can drive more than one transformation, so this experiment cannot attribute coverage or bug yield to a particular transformation. A finer-grained, per-handler ablation would require re-running the campaign with each active handler individually disabled, which we leave as future work (\S\ref{sec:discussion}).
\endgroup

\subsection{Parallelism reduces marginal coverage}
\label{subsec:parallel}

We next investigated whether running six tests concurrently exposes additional behavior.
We chose six instances because a typical consumer system has at least four CPU cores, making six concurrent browser instances a plausible stress configuration for a commodity testing machine.
We study only invalid mode for this comparison because it was more effective in \S\ref{subsec:singleanalysis}.
For each invalid scale we aggregated the logs from six parallel runs and compared them to a single run. Figure~\ref{fig:parallel-unique} presents coverage growth; Figure~\ref{fig:parallel-logs} shows cumulative log volume; Figure~\ref{fig:parallel-ttc} shows time-to-coverage milestones; and Figure~\ref{fig:parallel-final} shows the final unique counts.

Contrary to intuition, parallel runs consistently discover \emph{fewer} unique locations than single runs. At scale~80, the single run reaches 195 unique sites, whereas the parallel run reaches 131. The cumulative log volume is also lower in the parallel case. Parallel runs generally reach their own milestones more slowly, with exceptions such as the 50\% milestone at scale~40. Because each milestone is defined relative to its run's final total, these times do not compare a common absolute target.
Resource contention among six instances may reduce work completed within each query's timeout. The observed counts show that this concurrent configuration did not improve the aggregate logging footprint.

\subsection{Cross‑Version Comparison}
\label{subsec:version-comparison}

WebGPU implementations evolved rapidly during our study. To understand how implementation robustness changed over time, we repeated the single-run experiments on Chromium~133, released on February 4, 2025, roughly a year before Chromium~146, which was pre-release at the time of our experiments. Our goal is to compare the breadth of behaviors exercised in the two versions and examine how the distribution of log severities shifted.

\begin{figure}[t]
    \centering
    \includegraphics[width=\linewidth]{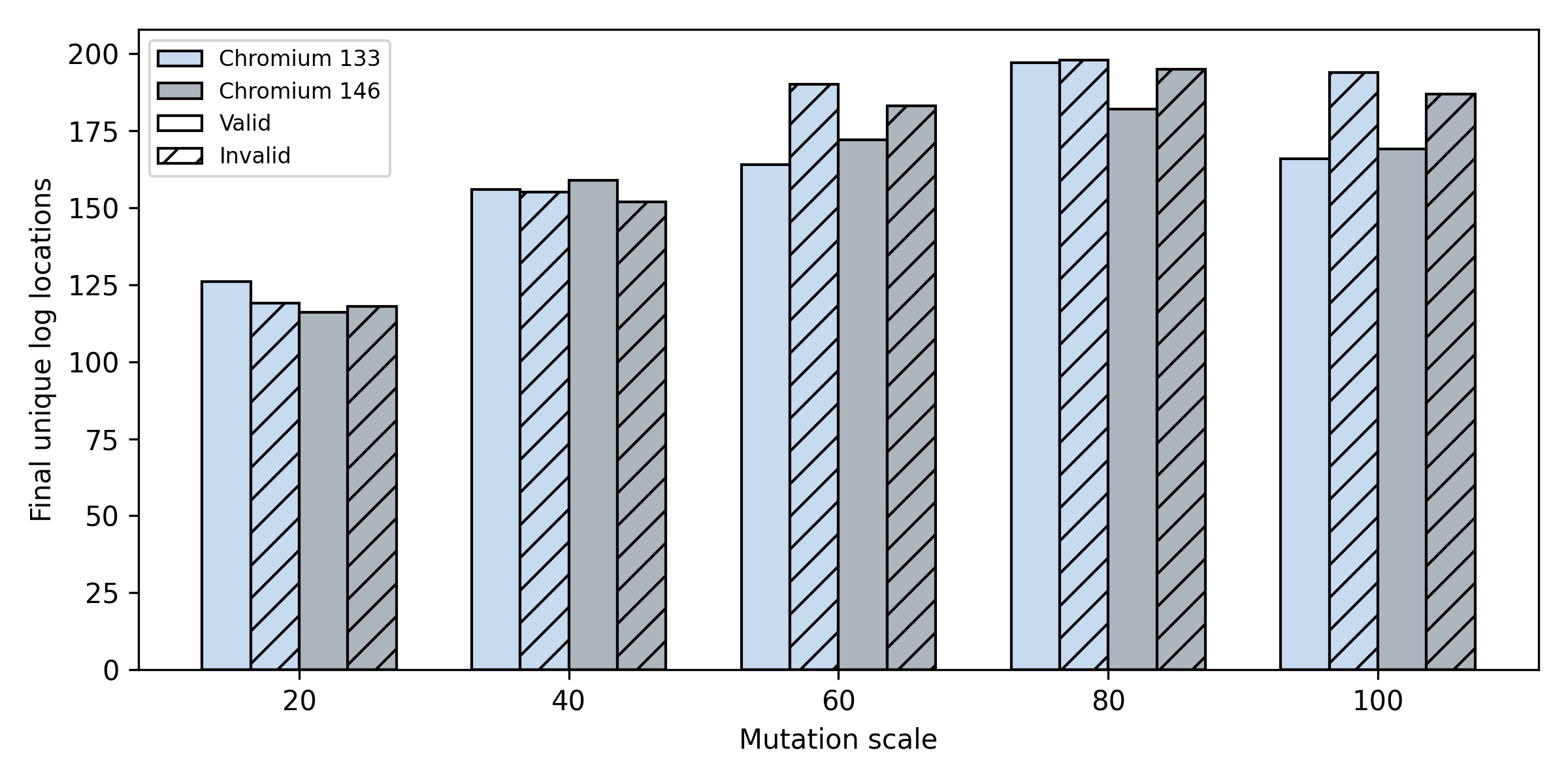}
    \caption{Final number of unique log‑emitting locations by mutation scale, version and mode.}
    \label{fig:version-final}
\end{figure}

\textbf{Coverage comparison.} Figure~\ref{fig:version-final} compares the total number of unique log-emitting locations for each mutation scale. Each scale has four bars: Chromium~133 in valid and invalid modes, and Chromium~146 in valid and invalid modes, which show similar trends across the versions for each mutation mode.

\begin{figure}[t]
    \centering
    \includegraphics[width=\linewidth]{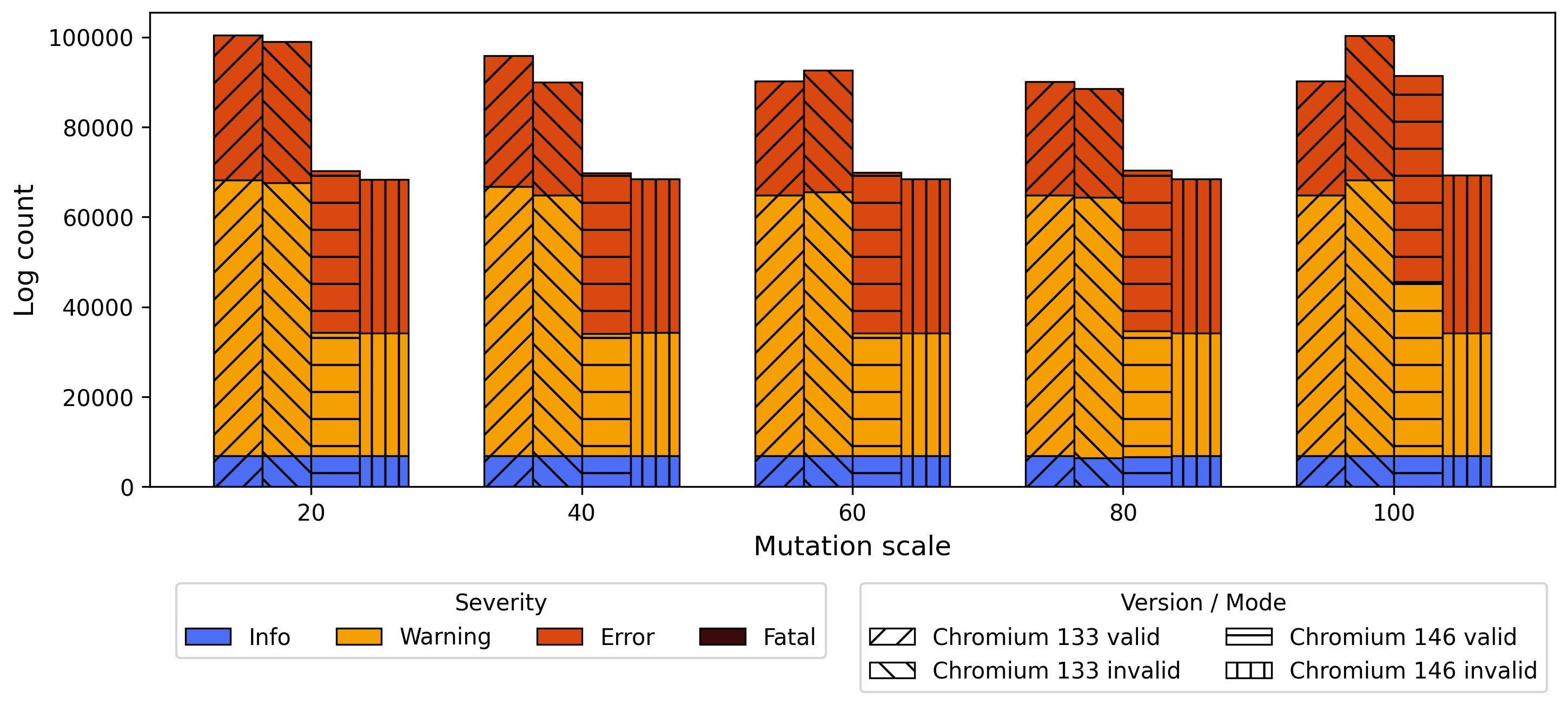}
    \caption{Severity distribution by mutation scale, version and mode. At each scale four bars represent Chromium~133 and~146 in valid and invalid modes; each bar is stacked to show the number of log entries at different levels.  Colors correspond to severity levels and hatching distinguishes valid and invalid modes.}
    \label{fig:version-severity}
\end{figure}

\textbf{Severity distribution.} Figure~\ref{fig:version-severity} aggregates the number of log entries for each severity (INFO, WARNING, and ERROR) across all scales.
First, Chromium~133 emits a higher number of log entries and a higher proportion of \emph{warnings} relative to \emph{errors}, whereas Chromium~146 produces a more balanced mix of warnings and errors. This shift could reflect changes in logging or executed behavior; the severity counts alone do not distinguish these explanations.
INFO logs remain a minority and no FATAL logs were observed in either version in this experiment.

\subsection{Findings}
\Sname recorded seven crashes (Table~\ref{tab:issues}). Among them, three findings are reproducible and we have responsibly reported them to Google. Based on the crash logs, we believe the non-reproducible ones may be due to race conditions or other nondeterministic behavior that we were unable to recreate.

\begin{table}[t]
  \centering
  \scriptsize
  \setlength{\tabcolsep}{4pt} \renewcommand{\arraystretch}{1.0}
  \resizebox{\columnwidth}{!}{\begin{tabular}{@{} l l c c @{}}
    \hline
    \textbf{Issue (short signature)} & \textbf{Location} & \textbf{Chr. Version} & \textbf{Rep.} \\
    \hline
    DCHECK\_font\_platform\_data & font\_cache\_skia.cc & 133 \& 146 & R \\
ASan\_Bad\_Free in V8 & thread\_isolated\_allocator.cc & 146 & R \\
    DCHECK\_put\_result & direct\_receiver.cc  & 133 & R \\
    DCHECK\_work\_areaIsEmpty & window\_sizer.cc & 133 & NR \\
    DCHECK\_data\_headernum\_bytes & backend\_impl.cc & 133 & NR \\
    DCHECK\_zero & file\_posix.cc & 133 & NR \\
    FATAL\_sandbox\_thread\_helpers & sandbox/thread\_helpers.cc & 146 & NR \\

    \hline
  \end{tabular}
  }
  \caption{Seven issues triggered by \Sname. Rep. denotes reproducibility (R = reproducible, NR = not reproducible); Chr. denotes Chromium.}
  \label{tab:issues}
\end{table}

\phantomsection\label{subsec:asan-case}\textbf{Case Study: V8 Heap Corruption.} Our fuzzing campaign uncovered a memory bug in Chromium~146 with ASan. The fuzzer generated an invalid test case at scale~80 whose mutation of the CTS triggered an invalid free (\texttt{attempting free on address which was not malloc()-ed}) during V8 (Chromium's JavaScript engine) garbage collection in its \texttt{ThreadIsolation} component. The browser attempts to unregister a just-in-time compilation (JIT) page
and invokes \texttt{PartitionAlloc} to free a pointer extracted from a corrupted memory allocation map (\texttt{std::\_\_Cr::\_\_tree<...>}). ASan detects that the pointer does not belong to a valid allocation and raises a bad-free exception. This bug occurs in V8's thread-isolation allocator, which is also responsible for enforcing write-xor-execute (W$\oplus$X) protections for JIT memory. Our test case reproduces reliably on both ASan-instrumented Chromium and on a non-ASan build. We reported this bug to the Chromium security team, who confirmed it as a duplicate of their non-public bug report. We contribute by providing a new triggering test case from WebGPU usage. This case study demonstrates that specification-guided mutation can uncover deep memory-safety violations in security-critical subsystems. Reaching this bug required preserving an extensive program state and ordering context, demonstrating the value of our semantics-aware mutation.

\phantomsection\label{subsec:mojo-case}\textbf{Case Study: Mojo Direct Receiver DCHECK Failure.} In Chromium~133, we triggered a DCHECK assertion failure in Mojo (Chromium's inter-process communication framework) while running a mutated CTS suite. The renderer aborted when \texttt{put\_result} was 5 (expected 0).
The stack shows that it is reached during Mojo receiver binding for Blink's widget input handler. The crash occurs on a Blink non-main thread. This renderer failure occurred during mutated CTS execution and is outside the WebGPU implementation. The stack alone does not establish a causal dependency on WebGPU calls or a security vulnerability; a DCHECK failure is a correctness signal requiring further triage.

\phantomsection\label{sec:eval:mode}\textbf{Mutation strategy lessons.}
Invalid-mode suites continue to discover new log-emitting locations late into a run (Figure~\ref{fig:invalid-coverage}), showing continued growth in the observed logging footprint. Moderate scales work best: scale~80 yields the highest final unique-location count, while scale~60 valid mode reaches high coverage especially quickly and scale~100 adds little coverage. An additional invalid-mode sweep on Chromium~146 (100 mutated suites per scale) matches this pattern: we observed 0 findings at scale~20, 3 at scale~40, 3 at scale~60, 4 at scale~80, and 1 at scale~100. Together with the parallelism results, this suggests that a single long-running invalid campaign around scale~80 is more effective than either aggressive parallelism or maximal-scale mutation.